\documentclass[12pt,a4paper]{article}
\usepackage{cite,amsmath,amssymb,bbm,color,graphicx}
\newcommand{\be}{\begin{equation}}  
\newcommand{\ee}{\end{equation}}  
\newcommand{\bea}{\begin{eqnarray}}  
\newcommand{\eea}{\end{eqnarray}}  
\newcommand{\ol}[1]{\overline{#1}}
\newcommand{\hc}{+\,\mathrm{h.c.}}
\newcommand{\vev}[1]{\langle #1 \rangle}

\newcommand\lsim{\mathrel{\rlap{\lower4pt\hbox{\hskip1pt$\sim$}}
    \raise1pt\hbox{$<$}}}
\newcommand\gsim{\mathrel{\rlap{\lower4pt\hbox{\hskip1pt$\sim$}}
    \raise1pt\hbox{$>$}}}

\newcommand{\tr}{\operatorname{tr}}

\providecommand{\tabularnewline}{\\}

\begin{document}

\vspace*{-2cm}
\begin{flushright}
  HD-THEP-09-10 \\
  LPSC 09-54 \\
  DCPT/09/74 \\
  IPPP/09/37 
\end{flushright}
\vspace*{1.2cm}

\begin{center}

{\LARGE\bf Phenomenology of Supersymmetric \\[2mm] 
Gauge-Higgs Unification}\\[12mm]

{\large F.~Br\"ummer$^{\,a}$, S.~Fichet$^{\,b}$, A.~Hebecker$^{\,c}$, 
S.~Kraml$^{\,b}$}\\[6mm]

{\it
$^{a}$~Institute for Particle Physics Phenomenology, Durham University,\\ 
Durham DH1 3LE, UK\\[3mm]
$^{b}$~Laboratoire de Physique Subatomique et de Cosmologie, UJF Grenoble 1, 
CNRS/IN2P3, 53 Avenue des Martyrs, F-38026 Grenoble, France\\[3mm]
$^{c}$ Institut f\"ur Theoretische Physik, Universit\"at Heidelberg, 
Philosophenweg 16 und 19, D-69120 Heidelberg, Germany}

\vspace*{12mm}

\begin{abstract}
\noindent Supersymmetric Gauge-Higgs Unification is a well-motivated new 
physics scenario, both in heterotic model building and from the perspective 
of higher-dimensional Grand Unified Theories. When combined with radion
mediated supersymmetry breaking, it allows for very specific predictions 
concerning the high-scale parameters of the MSSM. Using an appropriately 
modified version of a standard RGE evolution code (SuSpect), we derive 
low-scale predictions which can be tested at the LHC. The phenomenological 
success of our setting depends crucially on the 5d Chern--Simons term, which 
has not been used in previous, less encouraging studies of gauge--Higgs 
unification in supersymmetry. 
\end{abstract}

\end{center}

\clearpage

\section{Introduction}

If the LHC discovers supersymmetry, it will be a great challenge to 
relate the measured electroweak-scale parameters of the supersymmetric
Standard Model to more fundamental structures at a high energy scale. One 
of the best-motivated physics proposals in this context is supersymmetric 
grand unification~\cite{Georgi:1974sy,Dimopoulos:1981yj}. However, SUSY 
grand unification by itself places only rather limited constraints on the 
low-energy parameter space.

An elegant and natural further idea making this setting more predictive is 
gauge--Higgs unification (GHU). In models of this type, some or all of the Higgs 
scalars are extra-dimensional components of gauge fields.\footnote{In 
fact, GHU has a long history in non-supersymmetric models without grand 
unification (see e.g.~\cite{Fairlie:1979at} and many subsequent papers). 
Here, we take the SUSY GUT idea as our main paradigm.
} 
GHU is for instance realized in many grand-unified models derived from 
heterotic string theory, where one or both of the MSSM Higgs doublets can 
come from the untwisted sector (see e.g.~\cite{Buchmuller:2005jr}; for a 
recent review see~\cite{Nilles:2008gq}). At a simpler level, GHU can be 
realized in purely field-theoretic 5d or 6d orbifold GUT models~\cite{
Kawamura:2000ev}. These can be viewed as effective unified field theories, 
valid directly below the heterotic string scale. Such constructions 
receive independent support from the string-scale/GUT-scale 
problem~\cite{Kaplunovsky:1985yy} as follows: One of the possibilities for 
overcoming this problem is the compactification on anisotropic 
orbifolds~\cite{Ibanez:1992hc,Witten:1996mz,Hebecker:2004ce,Dundee:2008ts}, 
where one or two of the compactification radii are much larger than the 
string length scale. This naturally allows for an intermediate effective 
description in terms of a 5d or 6d orbifold GUT.

The present paper, which is mainly phenomenologically oriented, does not 
depend on specific string-theoretic realizations of GHU. Our analysis relies 
only on simple SUSY field theory models in which GHU can arise, and whose
4d low-energy effective field theory is the MSSM. The earliest 
and simplest construction of this type is the 5d SU(6) model of Burdman and 
Nomura~\cite{Burdman:2002se}, which we will largely follow. We expect, 
however, that our phenomenological results will carry over to similar
models, including more elaborate string-derived constructions.
Related models include, e.g., the 5d SU(6) model with warped extra dimensions 
of \cite{Nomura:2006pn} and the 6d models of \cite{Lee:2003mc,Buchmuller:2007qf,
Hosteins:2009xk}. For more work on SUSY GHU in orbifold GUTs, see for instance
\cite{Haba:2002vc} and references therein.

The main point of 5d GHU models relevant to our work is easily explained:
5d gauge symmetry enforces a K\"ahler potential in the Higgs sector of the 
form~\cite{Choi:2003kq} (see also~\cite{LopesCardoso:1994is,Brignole:1995fb,
Antoniadis:1994hg})
\be
S\supset\int d^4x\int d^4\theta\,\,\varphi\ol\varphi\,Y_H(T,\ol{T})\,\,
(\ol{H}_1+H_2)(\ol{H}_2+H_1)\,.\label{basic}
\ee
Here $H_1$ and $H_2$ are the MSSM Higgs superfields, arising from the adjoint
of the 5d gauge group after its breaking to the Standard Model group. The 
prefactor $Y_H$ is a real function of the radion superfield $T$, and $\varphi$
is the chiral compensator of 4d supergravity. If $T$ and $\varphi$ develop
$F$-term VEVs~\cite{Kobayashi:2000ak}, Eq.~(\ref{basic}) clearly induces a 
supersymmetric $\mu$ term as well as soft Higgs mass terms. They satisfy
the relation 
\be
m_1^2=m_2^2=|m_3^2|\label{ghubc}
\ee
at the GUT scale, where $m_{1,2}^2=|\mu|^2+m_{H_{1,2}}^2$ are the diagonal 
entries of the Higgs mass matrix and $m_3^2=B\mu$ is the off-diagonal 
element. This is a specific realization of the Giudice-Masiero 
mechanism~\cite{Giudice:1988yz} or of its string-theoretic 
version~\cite{Kaplunovsky:1993rd}. 

At the TeV scale, the familiar conditions for electroweak symmetry 
breaking and vacuum stability read:
\be
\begin{split}
m_1^2 m_2^2-m_3^4&<0,\\ 
m_1^2+m_2^2-2m_3^2&>0.\label{ewsbconditions}
\end{split}
\ee
Renormalization group (RG) running must thus turn the equalities of 
\eqref{ghubc} at the UV scale into the inequalities of 
\eqref{ewsbconditions} at the weak scale. A numerical analysis is 
required in order to find out whether this is possible at all and, if so, 
whether it is possible within a realistic model. This analysis has to take 
into account some additional predictions of 5d GHU models. In particular, 
there are strict relations between the Higgs mass parameters, the gaugino 
mass, and the dominant soft sfermion masses and trilinear couplings.

If the function $Y_H$ comes entirely from the gauge-kinetic term of the 5d 
super Yang--Mills action, one finds $Y_H\sim 1/(T+\ol{T})$. In this case, the 
relations between the soft parameters turn out to be rather restrictive 
and a realistic low-energy spectrum cannot be obtained without extreme 
fine-tuning~\cite{Choi:2003kq}. 

If we also include a supersymmetric 5d Chern--Simons term and allow for a 
VEV of the chiral adjoint in the 5d gauge multiplet, 
an extra contribution $\sim 1/(T+\ol{T})^2$ to 
the function $Y_H(T,\ol{T})$ arises~\cite{Hebecker:2008rk}. The 
Chern--Simons term is generically present in a 5d super Yang--Mills theory 
compactified on $S^1/\mathbb{Z}_2$, and its coefficient is determined by 
anomaly cancellation \cite{ArkaniHamed:2001is, Scrucca:2004jn}. It depends 
on the full field content of the model, which we do not fix completely. In 
particular, the fields of the two light generations can be distributed in 
various ways between bulk and branes. The prefactor of the 
$1/(T+\ol{T})^2$--contribution to $Y_H$ depends on the coefficient of the 
Chern--Simons term and on the size of the adjoint VEV, which is also unknown. 
Hence we treat this prefactor as an extra parameter.

The main point of the present paper is to demonstrate that, allowing for
this Chern--Simons term contribution to Eq.~(\ref{basic}), a completely 
realistic low-energy phenomenology can be obtained. This scenario is rather 
constrained since all MSSM soft parameters are given in terms of the VEVs 
of $F^T$ and $F^\varphi$, a dimensionless parameter $c'$ characterizing the 
effect of the Chern--Simons term, and two mixing angles related to the 
5d-origin of the third-generation quarks and leptons. We analyze the 
low-energy phenomenology of this setting and discuss observational 
consequences for the LHC.

To this end, we numerically solve the renormalization group evolution of the 
MSSM parameters, with GUT-scale boundary conditions as determined by our
GHU model. This procedure has been implemented within the public MSSM spectrum 
generator code {\tt Suspect\,2.41}~\cite{Djouadi:2002ze}. We perform two
extensive parameter scans. The first one uses a rather crude estimate for 
boundary conditions in the sfermion sector. This is essentially a 
generalization of the analysis of~\cite{Choi:2003kq}, now including the 
effects of a Chern--Simons term and using a state-of-the-art RG code. The 
second scan uses realistic boundary conditions, derived from the 
Burdman--Nomura model~\cite{Burdman:2002se}. It is somewhat more involved 
because of the specific relations between 4d Yukawa couplings and fundamental 
model parameters.

In both cases we find regions of parameter space where all present 
experimental bounds are satisfied. In particular, we can have sufficiently 
high Higgs and sparticle masses to evade direct search bounds, comply with 
$B$-physics constraints on rare decays, and also obtain a neutralino as the 
lightest supersymmetric particle (LSP) with 
a dark matter relic density compatible with WMAP results. We conclude that 
SUSY gauge--Higgs unification with radion mediated SUSY breaking is indeed 
a phenomenologically viable scenario. We also point out some characteristic 
LHC signatures which seem to be typical for this framework.

The paper is organized as follows: In Section \ref{sec:ghurelation}, we 
explain the appearance of the GHU relation, Eq.~\eqref{ghubc}, in a large 
class of models. In Section \ref{sec:5dmodel}, we present a concrete 5d 
model and give the formulas for the GUT-scale gauge- and Higgs-sector soft 
terms as functions of the model parameters. Section \ref{sec:running} 
contains a discussion of the expected RG running behaviour, particularly
of $m_3^2$, and of its consequences for identifying phenomenologically
promising parameter space regions. We discuss our general setup for 
numerically analyzing the RGEs in Section \ref{sec:generalnum}. Assuming a 
simplified set of sfermion soft terms, we present a first such analysis in 
Section \ref{sec:simplenum}. We proceed in Section \ref{sec:realistic} by  
explaining how a fully realistic sfermion sector can be included in the 
analysis. The corresponding numerical results are presented in 
Section \ref{sec:bnnum} and Conclusions are given in Section 
\ref{sec:conclusions}.

\section{The origin of the GHU relation for Higgs mass parameters}\label{sec:ghurelation}

In string-derived or orbifold GUT models, the MSSM Lagrangian will generically 
depend on several moduli fields. We focus on models where SUSY breaking is 
communicated to the MSSM predominantly through the moduli. Furthermore, we
assume that the K\"ahler potential depends on the Higgs superfields
$H_1$ and $H_2$ only in the combination $\ol{H}_1+H_2$ and $H_1+\ol{H}_2$. 
This assumption will be justified momentarily for a certain class of models. 
The leading part of the Higgs action then reads
\be\label{h1h2action}
S\supset\int d^4x\int d^4\theta\;\ol\varphi\varphi\;
Y_H( Z^I,\overline{Z}^{\bar J})\;(\ol H_1+H_2)(\overline{H}_2+H_1)\,,
\ee
where $Y_H$ is some real analytic function of the moduli fields $Z^I$. 

In GHU models there is no superpotential contribution
to the $\mu$ term. The Higgs mass parameters are therefore entirely determined
by Eq.~\eqref{h1h2action}: For canonically normalized Higgs 
fields they are given by 
\be\begin{split}\label{higgsmasses}
m_{H_1}^2&=m_{H_2}^2=-F^I\overline{F}^{\bar J}\partial_I\partial_{\bar J}
\log Y_H,\\
\pm\mu&=\overline{F}^{\bar\varphi}+\overline{F}^{\bar I}\partial_{\bar I}
\log Y_H,\\
\pm B\mu&=\left|F^\varphi+F^I\partial_I\log Y_H\right|^2-F^I
\overline{F}^{\bar J}\partial_I\partial_{\bar J}\log Y_H.
\end{split}
\ee
These equations obviously imply Eq.~\eqref{ghubc}.\footnote{
Note 
the sign ambiguity in $\mu$ and $B\mu$: Simultaneously changing the signs of 
both $\mu$ and $B\mu$ corresponds to a redefinition of one of the Higgs 
fields, say $H_1$, by $H_1\rightarrow -H_1$. The signs of the Yukawa couplings 
can be kept unchanged by an analogous redefinition of the right-handed matter 
fields it couples to. The overall sign of $\mu$ and $B\mu$ cancels in the 
RGEs. Therefore, given a certain high-scale model, a simultaneous sign change 
in the last two lines of Eq.~(\ref{higgsmasses}) will leave the absolute 
values of all weak-scale parameters unchanged, flipping only the signs of 
$\mu$ and $B\mu$ at the weak scale. The convention that $B\mu$ is positive at 
the weak scale then determines the signs in Eq.~(\ref{higgsmasses}).}

If we start from a 5d model, it is straightforward to see why the K\"ahler 
potential always depends only on the combination $H_1+\ol{H}_2$ (and its 
complex conjugate): Recall that a generic 5d super--Yang--Mills theory 
contains, in terms of 4d superfields~\cite{ArkaniHamed:2001tb,Marti:2001iw,
Hebecker:2001ke}, a vector superfield and a chiral adjoint $\Phi$. The fifth 
component of the gauge field, $A_5$, forms the imaginary part of the scalar 
component of $\Phi$. In the 5d kinetic action, the chiral adjoint appears 
only in the combination $\Phi+\Phi^\dagger$~\cite{Hebecker:2001ke}. The 
reason is that, in this combination, $A_5$ drops out of the lowest component 
of the real superfield $\Phi+\Phi^\dagger$, ensuring the absence of 
non-derivative 
couplings of $A_5$.\footnote{In 
fact, an analogous argument forces the K\"ahler potential to depend on the 
radion modulus $T$ only through the combination $T+\ol{T}$. In this case, 
the imaginary part is the 5th component of the graviphoton of 5d supergravity, 
and the real combination $T+\ol{T}$ ensures the absence of non-derivative 
couplings of this gauge-field component~\cite{Marti:2001iw}.}

The decomposition of the adjoint of the 5d gauge group $G$ 
in irreducible Standard Model 
representations is vector-like and can contain pairs of weak doublets: 
\be
{\bf Ad}(G)\,\rightarrow ({\bf 1}, {\bf 2})_{-1/2}+({\bf 1}, {\bf 2})_{1/2}
+\ldots\label{ad}
\ee
If this is the case, and if the zero-modes of these doublets in $\Phi$ survive 
compactification and GUT symmetry breaking, they can be identified with the 
Higgs fields $H_1$ and $H_2$. It is then only the combination $H_1+\ol{H}_2$ 
which appears in the 4d K\"ahler potential.

Intriguingly, this peculiar combination of chiral superfields has also been 
found in heterotic $E_8\times E_8$ orbifold models, where no use of an 
intermediate 5d effective theory is made~\cite{LopesCardoso:1994is,
Antoniadis:1994hg,Brignole:1995fb}. We briefly describe the situation 
following~\cite{Antoniadis:1994hg}: The K\"ahler potential can be 
expanded as
\be
K={\cal K} + {\cal Y}_{\alpha\bar\alpha} A^\alpha \bar A^{\bar\alpha}
+\tilde{\cal Y}_{\beta\bar\beta} B^\beta \bar B^{\bar\beta}+
\left({\cal Z}_{\alpha\beta}A^\alpha B^\beta\hc\right)+\ldots
\ee
Here ${\cal K},{\cal Y}, \tilde{\cal Y}, {\cal Z}$ are functions of the 
moduli, and $A^\alpha$ and $B^\beta$ are matter fields transforming in the 
${\bf 27}$ and $\ol{\bf 27}$ of $E_6$, respectively. Suppose that the MSSM 
Higgs fields $H_1$ and $H_2$ are contained in two such fields $A$ and $B$. 
For the desired combination $H_1+\ol{H}_2$ to appear, one needs
\be
{\cal Y}_{A\bar A}=\tilde{\cal Y}_{B\bar B}={\cal Z}_{AB}=
\ol{\cal Z}_{\bar A\bar B}\,.\\ \label{YYZ}
\ee
Indeed, for untwisted matter fields $A,B$ associated with a common complex 
plane, it was found that
\be
{\cal Y}_{A\bar A}=\tilde {\cal Y}_{B\bar B}={\cal Z}_{AB}=
\ol{\cal Z}_{\bar A\bar B}=\frac{1}{(T+\ol T)(U+\ol U)}\,,
\ee
where $U$ and $T$ are the complex structure and K\"ahler modulus of that 
plane. By contrast, for twisted matter fields, or for untwisted matter fields 
associated with distinct planes, or with a plane without $U$ modulus, one 
has ${\cal Z}_{AB}=0$.

We now argue that the above string-theoretic results are closely related to
the previous 5d argument. Firstly, if $A$ and $B$ are associated with the same 
complex plane, one can take a 6d limit in which the two corresponding 
compactification radii remain large. In this limit, the fields $A$ and $B$ 
are the extra-dimensional components of the 6d gauge field. Secondly, the 
presence of a shape modulus $U$ of the corresponding large $T^2$ allow for 
5d limit, in which the compact space becomes an interval. Thus, the 
string-theoretic conditions which ensure that the combination $H_1+\ol{H}_2$ 
appears are precisely those which are needed for an appropriate 5d limit to 
exist. Even in regions of moduli space which do not correspond to that 
limit, the structure enforced by 5d gauge invariance survives. In other 
words, the simple 5d argument given earlier appears to be sufficient to 
understand the situation in heterotic orbifold models as well. It would be 
interesting to work this out in more detail, which is however beyond the 
scope of the present, mainly phenomenologically oriented paper. 

As we have already emphasized in the Introduction, and as will become clear 
in subsequent sections, the 5d supersymmetric Chern--Simons term plays a 
central role in our analysis. Such a term cannot arise in the tree-level 
dimensional reduction of a theory with more than five dimensions. However, 
it is generically induced at one loop~\cite{Seiberg:1996bd}. More 
specifically, the radiative generation of a Chern--Simons term in 
compactifications from 6d to 5d was discussed in~\cite{Hebecker:2004xx}. 
Thus, the 5d Chern--Simons term is consistent with a 10d origin of the 
theory.

\section{A concrete 5d realization}\label{sec:5dmodel}

We now turn to the phenomenological prospects of concrete 5d models. Our 
main example will be a generalization of the SU$(6)$ model of Burdman and
Nomura~\cite{Burdman:2002se}, with a larger gauge group containing at least 
U$(6)=$ SU$(6)\times$U$(1)$. The 5d gauge theory is compactified on 
$S^1/(\mathbb{Z}_2\times\mathbb{Z}_2')$. The only relevant modulus is the 
radion superfield $T=\rho+iB_5+\ldots\,$, where $B_5$ is the fifth component 
of the graviphoton. The real part $\rho$ is normalized such that $2\pi\rho$ 
is the volume of the original $S^1$. We assume that it is eventually 
stabilized at $\langle \rho\rangle = R$. The Higgs doublets are contained 
in the superfield $\Phi=\Sigma+iA_5+\ldots$ (where $\Sigma$ is the chiral
adjoint of the 5d gauge multiplet, and $A_5$ is the fifth component of the 
gauge boson). 

The U$(1)$ gauge factor is assumed to be broken on the boundary. Furthermore, 
orbifold boundary conditions for the gauge fields can be chosen such
\cite{Burdman:2002se} that the remaining SU$(6)$ is broken in 4d to 
the Standard Model,\footnote{
Apart 
from an extra U$(1)$ which is Higgsed by a brane field.
} 
and that the only components of $\Phi$ with zero modes are the Higgs 
fields. Their origin is particularly obvious in the SU(5)-decomposition 
\be
{\bf 35}={\bf 24}+{\bf 5}+\overline{\bf 5}+{\bf 1}
\ee
of the adjoint of SU(6). Our fields $H_1$ and $H_2$ are the doublets 
contained in the $\overline{\bf 5}$ and ${\bf 5}$ respectively. 

Before orbifolding, the 4d effective theory contains the full gauge multiplet
and chiral adjoint $\Phi$ as well as the radion $T$. The corresponding 
leading-order action, which has been analyzed in \cite{Marti:2001iw}, 
contains a term 
\be
S\supset \frac{\pi R}{g_5^2}\int d^4x\,\int d^4\theta\,\ol\varphi\varphi\,
\frac{2R}{T+\ol T}\tr\left(\Phi+\Phi^\dag\right)^2\,.
\ee
Retaining only the $\Phi$ components that survive the orbifold projection, 
i.e.~the Higgs fields, this becomes~\cite{Choi:2003kq}
\be\label{s4gk}
S\supset \frac{2\pi R}{g_5^2}\int d^4x\int d^4\theta\,\ol\varphi\varphi\,
\frac{2R}{T+\overline{T}}(\ol H_1+H_2)(\ol H_2+H_1).
\ee
Note that the coupling of $T$ to the Higgs fields depends on the choice 
of K\"ahler--Weyl frame. We work in a frame where $K=-3\log(T+\ol T+\ldots)$. 
The 4d metric has not been rescaled after compactification. 

The 5d theory will in general also contain a Chern--Simons term. Its
supersymmetrized version includes a cubic term in $\Phi$, which couples to 
the radion according to~\cite{Hebecker:2008rk}
\be
S\supset\frac{c\,\pi R}{3}\int d^4x\int d^4\theta\,\ol\varphi\varphi\,
\left(\frac{2R}{T+\ol T}\right)^2
\tr\left(\Phi+\ol\Phi\right)^3.
\ee
After orbifolding and allowing for a non-zero expectation value 
$\vev{\Phi}=v\,\mathbbm{1}_6$, this part of the Chern--Simons action 
contributes as 
\be\label{s4cs}
S\supset\frac{2c'\pi R}{g_5^2}\int d^4x\int d^4\theta\,\ol\varphi\varphi\,
\left(\frac{2R}{T+\overline{T}}\right)^2\,(\ol H_1+H_2)(\ol H_2+H_1)
\ee
to the quadratic Higgs Lagrangian.\footnote{Our 
$\Phi$-VEV $v=v_{4d}$ is a VEV in the 4d effective theory. It is 
related to the corresponding VEV $v_{5d}$ of the underlying 5d theory by 
$v_{4d}=(\rho/R)v_{5d}$. Here the mass dimensions of the 4d and 5d scalar
fields are the same since we assume that the complete leading-order 5d 
Lagrangian has a prefactor $1/g_5^2$.
}
Here we have introduced the dimensionless 
constant $c'=2cv g_5^2$. Note that the group \mbox{U$(6)=$ SU$(6)
\times$U$(1)$}, which we use here and below, is only the simplest extension 
of SU(6) allowing for an SU(6)-preserving $\Phi$-VEV.\footnote{The 
significantly more complicated case of SU$(6)$-breaking expectation values 
is considered in \cite{Hebecker:2008rk}.
}
Larger and, in particular, simple groups, such as SU(7), are clearly possible.

We regard $c'$ as a free parameter for the purpose of our analysis: While
boundary-anomaly cancellation fixes $c$ for a given field content, we 
specify neither this field content nor the value of $v$. In particular, 
different distributions of the light generations between the two branes
affect the value of $c$. However, positivity of the kinetic terms (given
just below) leads to the constraint
\be
c'>-1\,.
\ee

The MSSM scalar potential for canonically normalized Higgs fields (which we 
also call $H_1$ and $H_2$ by abuse of notation) reads, at quadratic order,
\be
V=m_1^2|H_1|^2+m_2^2|H_2|^2+m_3^2(H_2 H_1+{\rm h.c.})\,.
\ee
The mass parameters $m_i^2$ can now be calculated from the Higgs kinetic 
function 
\be\label{YH}
Y_H(T,\overline{T})=\frac{\pi R}{g_5^2}\left(1+c'\frac{2R}{T+\overline{T}}
\right)\frac{2R}{T+\overline{T}},
\ee 
which follows from Eqs.~\eqref{s4gk} and \eqref{s4cs}. According to 
Eqs.~\eqref{higgsmasses}, the parameters $\mu$ and $m_i^2$ 
read~\cite{Hebecker:2008rk}
\be
\epsilon_{H}^{}\,\mu = \ol F^{\bar\varphi}-\frac{\ol F^{T}}{2R}\frac{1+2c'}{1+c'}\,,
\label{mubmu1}
\ee
\be\label{mubmu2}
  m_{1}^2 = m_{2}^2 = \epsilon_{H}^{} m_3^2 = 
  |F^\varphi|^2 - \frac{(F^\varphi\overline F^{T}+{\rm h.c.})}{2R}
  \frac{1+2c'}{1+c'} +  \frac{|F^T|^2}{(2R)^2}\frac{2{c'}^2}{(1+c')^2}\,,
\ee
where we have introduced a parameter $\epsilon_{H}^{}=\pm 1$ to 
account for the sign ambiguity in $m_3^2=B\mu$ and $\mu$.

From the 5d gauge-kinetic and Chern--Simons action we obtain, after dimensional
reduction,
\be
S\supset\frac{\pi R}{g_5^2}\int d^4x\int d^2\theta\,\frac{T}{R}\tr 
W^\alpha W_\alpha\hc + 2c\pi R\int d^4x\int d^2\theta\,\tr\left(\Phi\, 
W^\alpha W_\alpha\right)\hc\,,
\ee
which eventually gives the 4d gauge-kinetic term
\be
S\supset \frac{\pi R}{g_5^2}\int d^4x\int d^2\theta\,\left(\frac{T}{R}+
c'\right)\,\tr W^\alpha W_\alpha\,\hc\,.
\ee
It determines the gaugino masses~\footnote{
Note 
that this agrees with~\cite{Hebecker:2008rk} only after a substitution
$c'\to c'/2$, which is due to our modified definition of $c'$. However, 
after this substitution, it becomes apparent that Eqs.~(\ref{mubmu1}) and 
(\ref{mubmu2}) are truly different from~\cite{Hebecker:2008rk}. This 
is the result of our SU(6)-preserving $\Phi$-VEV, as opposed to the 
SU(6)-breaking $\Phi$-VEV of~\cite{Hebecker:2008rk}.
}
\be\label{mhalf}
M_{1/2}=\frac{\overline{F}^T}{2R}\frac{1}{1+c'}\,
\ee
as well as the 4d gauge coupling
\be
\frac{1}{g_4^2}=\frac{2\pi R}{g_5^2}(1+c')\,.
\ee

The soft masses and trilinear terms for the  matter multiplets are more 
model-dependent. Quite generally the relevant piece of the kinetic action can 
be written as
\be
\begin{split}
S\supset\int d^4x\int d^4\theta\,\ol\varphi\varphi\,\Bigl[&Y_{U}
(T,\overline{T})\,|U|^2+Y_{Q}(T,\overline{T})|Q|^2+Y_{D}
(T,\ol T)|D|^2\\
&+Y_{E}(T,\overline{T})\,|E|^2+Y_{L}(T,\overline{T})|L|^2
+Y_{N}(T,\ol T)|N|^2\Bigr].
\end{split}
\ee
The kinetic functions $Y_{X}$ (with $X=U,D,Q,E,N,L$ standing for up-type 
and down-type right-handed quarks, quark doublets, charged and uncharged 
right-handed leptons and lepton doublets) determine the soft masses 
according to
\be\label{eq:mx}
m_{X}^2=-|F^T|^2\frac{\partial^2}{\partial T\partial\overline{T}} 
\log Y_{X}(T,\overline{T})\,.
\ee
The trilinear couplings are given by
\be\label{eq:aud}
A_{U,D}=F^T\frac{\partial}{\partial T}\log\left(Y_H Y_{Q}Y_{U,D}\right),
\ee
\be\label{eq:ae}
A_{E}=F^T\frac{\partial}{\partial T}\log\left(Y_H Y_{L}Y_{E}\right).
\ee
Note that we define the $A$ term with a negative sign in the Lagrangian:
\be
{\cal L}\supset -\left(A_U y_U H_2 \tilde Q\tilde U
+A_D y_D H_1 \tilde Q\tilde D+A_E y_E H_1 \tilde L\tilde E\hc\right).
\ee

The precise form of the matter kinetic functions depends on the model under 
consideration. We will assume throughout that the first two generations of
MSSM matter are brane-localized and that their GUT-scale soft terms are
negligible. This gives no-scale boundary conditions for the first and second 
generation. For the third generation, we will consider two cases: 
first the approximation that only the top quark receives a Yukawa coupling 
induced by the 5d gauge coupling, and second the case of the Burdman--Nomura 
model with realistic top, bottom and tau Yukawa couplings.

\section{Expected running patterns}\label{sec:running}

Before we present the numerical results, let us briefly discuss what general
features we expect. Needless to say, the complete system of two-loop 
renormalization group equations (RGEs), which will be solved numerically in 
the following sections, is far too complicated to permit an analytical 
treatment. Some aspects can nevertheless be qualitatively understood by 
inspection of the dominant contributions to the one-loop RGEs. 

The scale of electroweak symmetry breaking (EWSB) is defined as usual as the
geometric mean of the stop masses, $M_{\rm EWSB}=\sqrt{m_{\tilde t_1}
m_{\tilde t_2}}$. At this scale the conditions of Eq.~\eqref{ewsbconditions} 
have to hold. Once EWSB occurs, one finds the well-known relations 
between $\mu$, $B\mu=m_3^2$, $M_Z$, the Higgs soft masses $m_{H_i}^2$
and the ratio $\tan\beta$ of Higgs expectation values:
\be
  \begin{split}\label{musugra}
    \mu^2 &= \frac{1}{2}\left[ \tan 2\beta\,(m_{H_2}^2\tan\beta 
	- m_{H_1}^2\cot\beta) - M_Z^2 \right] , \\
    B\mu &=  \frac{1}{2} \sin 2\beta \left[ m_{H_1}^2 
	+ m_{H_2}^2 + 2\mu^2 \right] .
  \end{split}
\ee

We focus on the region of moderately large $\tan\beta$, roughly 
$\tan\beta\gtrsim 5$, to ensure that the tree-level bound on the lightest 
Higgs mass, $m_{h}\leq M_Z$, is approximately saturated. The Higgs mass 
can then be lifted above the direct search limit by radiative corrections, 
mainly due to stop loops. 

The latter involves a significant fine-tuning (the notorious MSSM ``little 
hierarchy problem''), because the soft mass scale must be large compared to 
$M_Z$ instead of being of the same order of magnitude, which would be the 
natural situation. For sizeable $\tan\beta$ one has
\be
\frac{M_Z^2}{2}\approx -m_2^2,
\ee
so $m_2^2$ must be negative and small compared to typical soft masses.
We will not discuss this fine-tuning any further (see however~\cite{
Barbieri:2000gf}), but accept it and focus on the implications for 
models with GHU boundary conditions. One immediate consequence is that 
$m_2^2>0$ at the GUT scale, because the soft mass $m_{H_2}^2$ and hence also 
$m_2^2$ runs down towards lower energies (the running of $\mu$ is 
insignificant). While this also fixes the GUT-scale sign of $m_1^2$ to be 
positive, 
either sign for $B\mu$ is possible  
(cf. Eq.~\ref{higgsmasses}). In other words, we can have $\epsilon_{H}=+1$ 
or $\epsilon_{H}=-1$ in the GHU relations
\be\label{ghubound}
  m_{H_1}^2 = m_{H_2}^2 = \epsilon_{H}^{} B\mu - |\mu|^2 \,.
\ee

However, as we will now argue, $\epsilon_{H}$ is always determined by the 
sign of $\mu$: Out of the four sign choices $\mu>0$ or $\mu<0$
and $\epsilon_H^{}=\pm 1$, only two can generically lead to realistic spectra. 
To establish this observe first that $m_1^2$ will typically not evolve by more 
than a factor of $2-3$, and therefore remains of the order of magnitude of the
typical soft mass scale during RG running. Furthermore, we just stated 
that $m_2^2$ at $M_{\rm EWSB}$ should be small compared to the typical soft
mass scale, and that $\tan\beta$ should at least be moderately 
large. From all this it follows that $B\mu$ at the EWSB scale should be small 
compared to the typical soft mass-squared scale as well, as can be read off from 
\be
\tan\beta+\cot\beta=\frac{m_1^2+m_2^2}{2 B\mu}
\ee
(which is equivalent to the second line of Eqs.~\eqref{musugra}). We will now
show that requiring small EWSB-scale $B\mu$ generically fixes $\epsilon_H^{}$ 
in terms of sign$(\mu)$.

The RG evolution of $B\mu$ is primarily governed by the terms involving the 
top trilinear coupling and the weak gaugino mass:
\be\label{Bmurge}
16\pi^2\frac{d}{dt}B\mu=\mu(6 A_t |y_t|^2+6 g_2^2 M_2)+\ldots
\ee
We can choose positive gaugino masses without loss of generality. Let us 
now discuss the relevance of the two dominant terms on the r.h. side of 
Eq.~(\ref{Bmurge}):

The gluino contribution to the $A_t$-RGE forces $A_t$ to run negative 
towards low scales. This is fairly universal, i.e.~more or less 
independent of the values of the other parameters. The value of $A_t$ at 
any given scale is thus to a good approximation dictated only by its 
GUT-scale boundary value and $M_{1/2}$. The running of gaugino masses is 
also approximately universal: at one-loop, they simply evolve according 
to the respective gauge coupling beta functions. In the RG evolution of $B\mu$, 
$A_t$ will therefore always dominate at low energies, when it has become 
large and negative and when also $y_t$ has grown large. Correspondingly, 
the $M_2$ term on the r.h. side of Eq.~(\ref{Bmurge}) can dominate only at 
energies near $M_{\rm GUT}$, before it is overwhelmed by $A_t$. 

For negative $\mu$, $B\mu$ initially increases from its GUT-scale value 
and then runs down; for positive $\mu$, it evolves in the opposite 
way. The relative importance of the $A_t$ and the $M_2$ contributions is set 
by their GUT-scale initial values: the larger $A_t$ at $M_{\rm GUT}$, the 
longer it will take to run negative and to finally dominate the $B\mu$ RG 
evolution. For small or negative GUT-scale $A_t$, the $B\mu$ running at low 
energies is more important than the initial, 
$M_2$-dominated phase near $M_{\rm GUT}$.

\begin{figure}[htb]\begin{tabular}{cc}
   \includegraphics[width=0.5\textwidth]{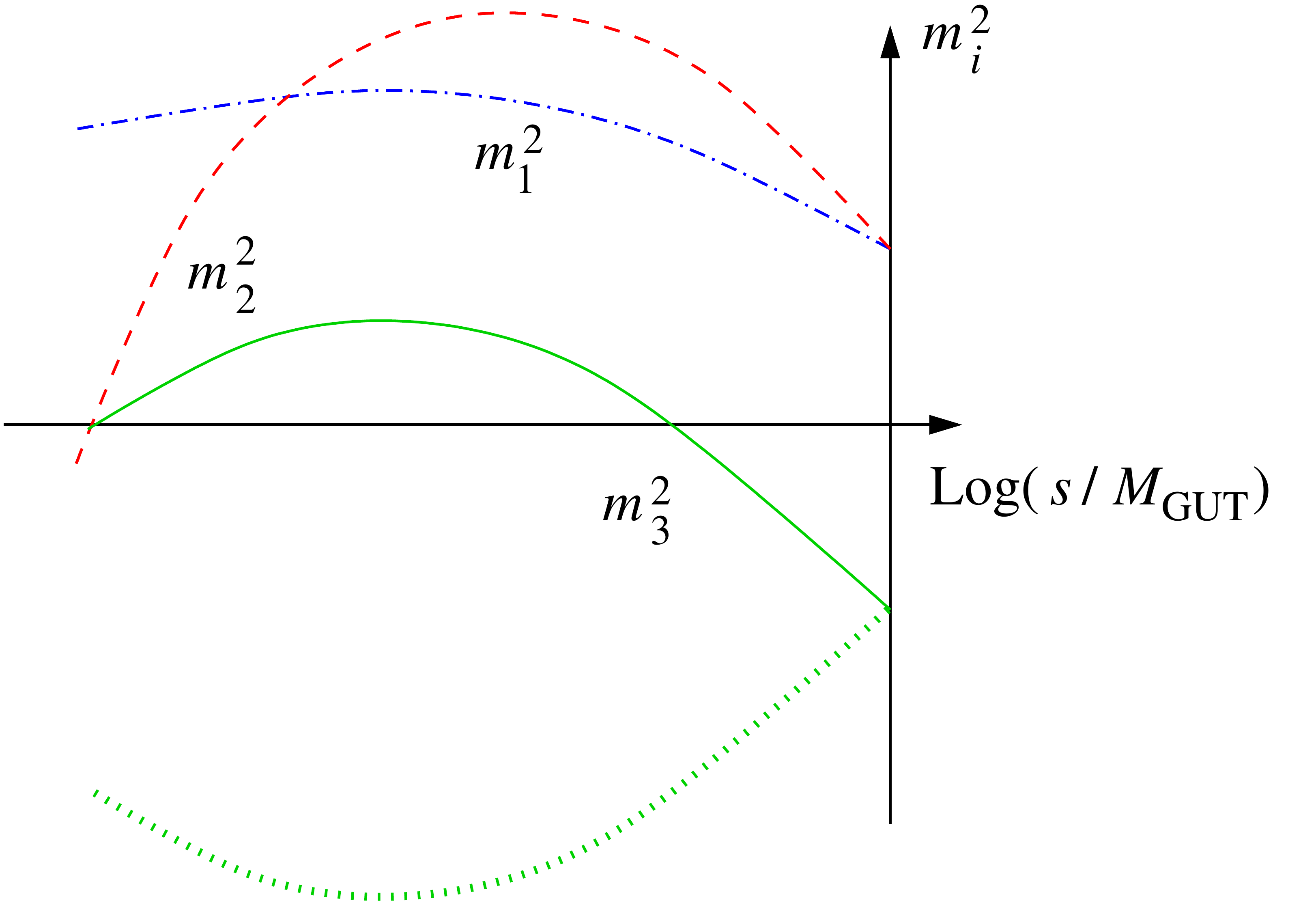}& 
   \includegraphics[width=0.5\textwidth]{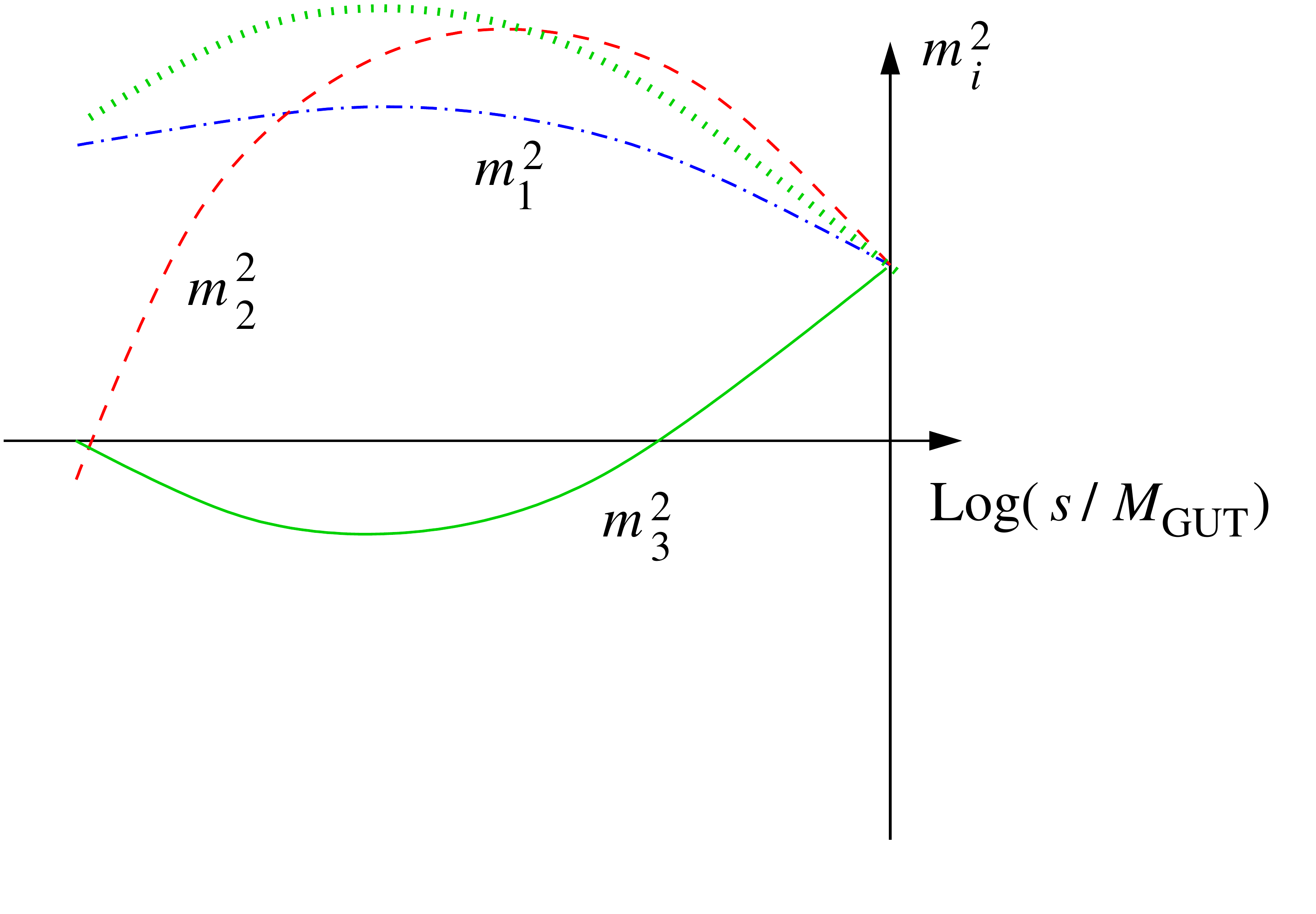}
   \end{tabular}
   \caption{Qualitative RG evolution of $m_1^2$ (blue dot-dashed curve), $m_2^2$ 
   (red dashed curve), and $m_3^2$ (green solid curve) as a function of the scale $s$,
   between $s=M_{\rm EWSB}$ and $s=M_{\rm GUT}$. 
   $A_t$ at the GUT scale is sizeable and positive. The green dotted curve is
   $m_3^2$ for the wrong sign($\mu$), which does not lead to realistic EWSB.
   Left panel: $\epsilon_H=-1$ requires sign$(\mu)=-1$. Right panel:
   $\epsilon_H=+1$ requires sign$(\mu)=+1$.}
   \label{fig:runningsketch1}
\end{figure}

\begin{figure}[htb]\begin{tabular}{cc}
   \includegraphics[width=0.5\textwidth]{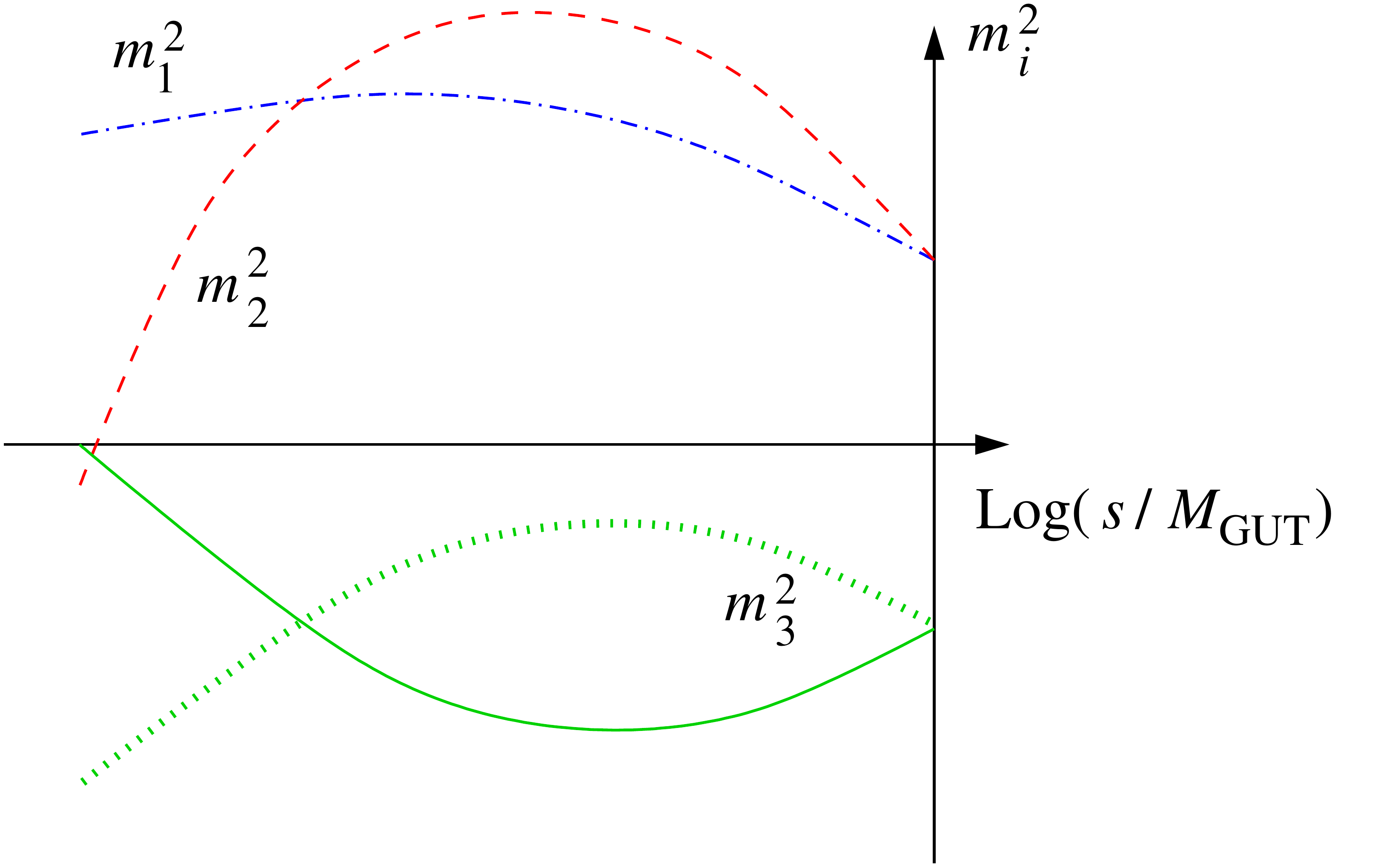}&
   \includegraphics[width=0.5\textwidth]{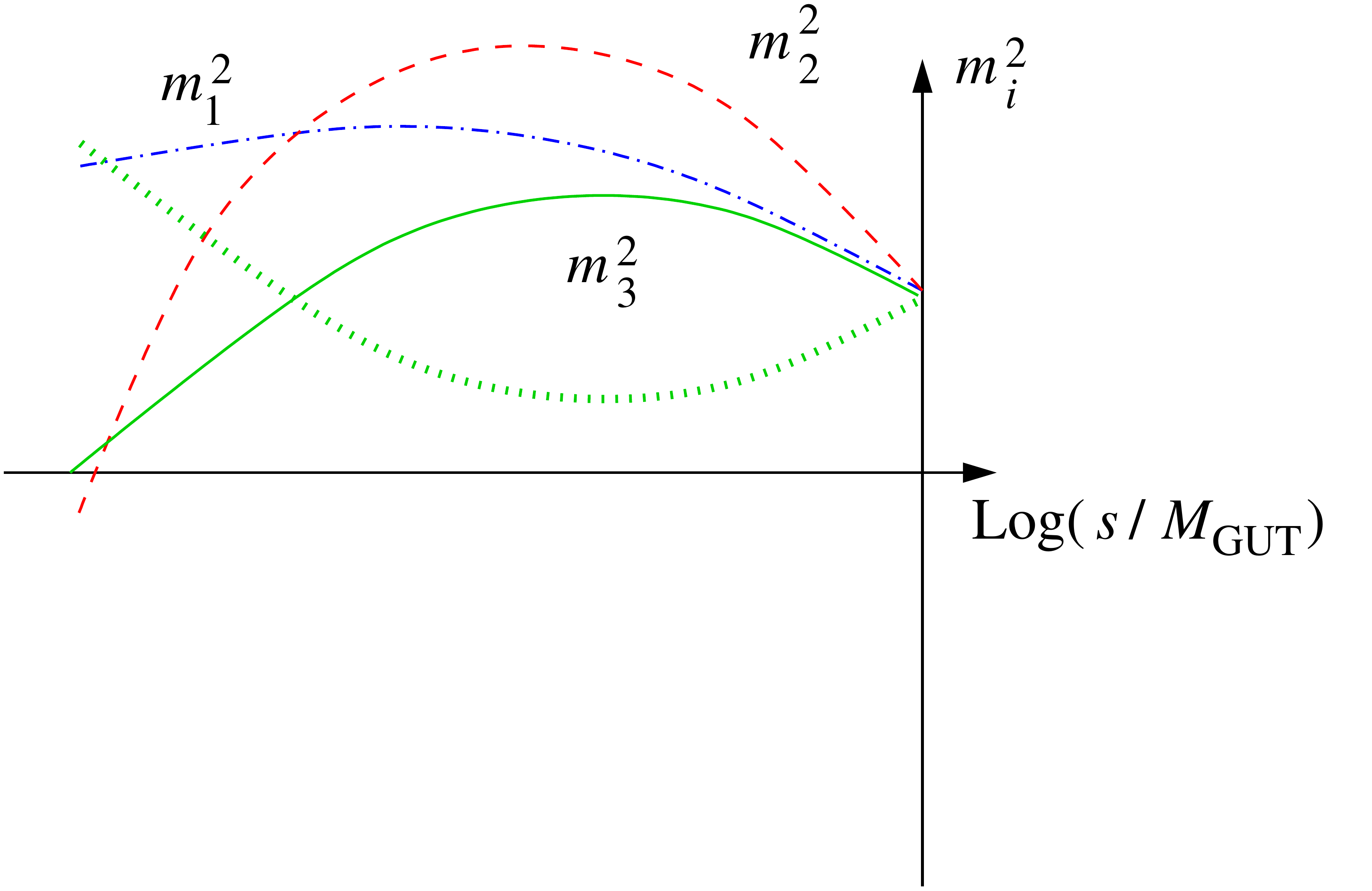}
   \end{tabular}
   \caption{Same as Fig.~\ref{fig:runningsketch1} for small or negative GUT-scale
   $A_t$.
   Left panel: $\epsilon_H=-1$ requires sign$(\mu)=+1$. Right panel:
   $\epsilon_H=+1$ requires sign$(\mu)=-1$.}
   \label{fig:runningsketch2}
\end{figure}

The direction and slope of the running of $B\mu$ are set by the sign and magnitude 
of $\mu$, which itself does not run significantly as mentioned. We observed before
that $B\mu$ should be small at the EWSB scale --- for the sake of the argument, let 
us try to construct a situation where it is exactly zero. It should in particular 
change significantly with respect to its initial GUT-scale value, so $|\mu|$ should 
be sizeable. Furthermore, changing the sign 
of $\mu$ will lead to $B\mu$ evolving in the opposite way (at least as far as the 
evolution is governed by the terms in Eq.~\eqref{Bmurge}). If there is a solution 
with, e.g., $\epsilon_{H}=-1$ and sign$(\mu)=-1$, we therefore expect a nearby mirror 
solution for the opposite sign choice. On the other hand, changing only one of
the signs will generically not lead to a solution due to the approximately 
universal behaviour of $A_t$ and $M_2$. We have sketched this behaviour in 
Fig.~\ref{fig:runningsketch1} for large GUT-scale $A_t$, and in 
Fig.~\ref{fig:runningsketch2} for small or negative GUT-scale $A_t$.

Thus, while one might naively have expected four branches of solutions of the RGEs 
to give realistic spectra (corresponding to the two choices of each sign$(\mu)$ 
and of $\epsilon_{H}$), by the above discussion there should in fact appear
only two. Furthermore, we expect sizeable $|\mu|$ in all cases.

\pagebreak
\section{Numerical analysis: General setup}\label{sec:generalnum}

For the numerical analysis, we make use of the public state-of-the-art SUSY 
spectrum code {\tt Suspect\,2.41}~\cite{Djouadi:2002ze}, appropriately modified
to be applied to our SUSY GHU model. The usual procedure in {\tt Suspect} and
in other SUSY spectrum codes is to use $\tan\beta$ and $M_Z$ as inputs and to 
compute $\mu$ and $B\mu$ from the EWSB condition Eqs.~\eqref{musugra}. In our 
model, however, the Higgs soft masses and $\mu$ and $B\mu$ are not independent, 
since they are related by the GHU conditions \eqref{ghubound} at the GUT scale. 

The free parameters in the gauge--Higgs sector of the model are actually
$F^T/2R$, $F^\varphi$ and $c'$, from which the GUT-scale values for $M_{1/2}$, 
$\mu$, $B\mu$ and $m_{H_{1,2}}^2$ are determined according to
 Eqs.~\eqref{mubmu1}, \eqref{mubmu2} and \eqref{mhalf}. Together with the 
GUT-scale values for the sfermion mass parameters and trilinear couplings, 
they furnish a set of GUT-scale boundary conditions for the MSSM 
renormalization group equations.

It is in principle possible to change the usual procedure of spectrum 
computation such that $\mu$ and $B\mu$ become high-scale inputs, while 
$\tan\beta$ as well as $M_Z$ are output determined by Eq.~\eqref{musugra}.
We have implemented this scheme in {\tt Suspect\,2.41}; the requirement 
to find the correct experimental value of $M_Z$, however, makes parameter 
scans very inefficient.

For the present analysis we have therefore chosen a different approach:
We work with the conventional SUGRA scheme of {\tt Suspect\,2.41}, which takes 
$\tan\beta(M_Z)$ together with the GUT-scale values of all soft-breaking 
parameters except $B\mu$ as input. We only modify this scheme by not specifying
fixed GUT-scale values for the Higgs soft masses $m_{H_1}^2$ and $m_{H_2}^2$,
but instead determining them from the GHU boundary conditions 
Eq.~\eqref{ghubound}.

Our input parameters are thus $M_{1/2}(M_{\rm GUT})$ and $\tan\beta(M_Z)$, 
the two sign coefficients ${\rm sign}(\mu)$ and $\epsilon_{H}^{}$,  
plus the sfermion mass parameters and $A$-terms at $M_{\rm GUT}$.
The values of $\mu$, $B\mu$, $m_{H_1}^2$, $m_{H_2}^2$ are computed 
iteratively applying Eqs.~\eqref{musugra} at the EWSB scale and 
Eq.~\eqref{ghubound} and the GUT scale. 
When a stable solution is found, the model parameters $F^T/2R$, $F^\varphi$ and $c'$ are  
inferred from $M_{1/2}$,  $\mu$ and $B\mu$ at  $M_{\rm GUT}$ by inverting 
Eqs.~\eqref{mubmu1}, \eqref{mubmu2} and \eqref{mhalf}. 

A complication arises, however, from the sfermion sector. 
As discussed in Section~\ref{sec:5dmodel},
we assume no-scale boundary conditions, i.e.~a common scalar mass 
$m_0\equiv 0$ and a common trilinear coupling $A_0\equiv 0$,  
for squarks and sleptons of the first two generations.
The soft terms of the third generation, on the other hand, can be non-zero.
To be more precise, they will depend on $F^T/2R$ and $c'$ 
(and possibly also on other model parameters)
according to their kinetic functions. 
This requires an extra level of iteration in the spectrum computation.

It turns out to be convenient to let this iteration act on $c'$. 
We thus start the procedure described above with an initial guess of $c'$, 
which is kept constant until a first convergence of the spectrum is reached. 
This has the virtue that the GUT-scale sfermion soft masses are unambiguously 
fixed in terms of $c'$, $M_{1/2}$, and other input parameters (as will become clear 
once we describe how we are modelling the matter sector) 
so the EWSB scale does not change too much in each iteration step, 
which could lead to numerical instabilities. When convergence is reached, 
an updated value of $c'$ as computed from $M_{1/2}$,  $\mu$ and $B\mu$ 
is taken as the new input $c'$, and the whole procedure is iterated until 
$c'$ converges as well.

Let us finally list the Standard Model (SM) input values and experimental constraints.
For the SM input values, 
we take $\alpha^{-1}(M_Z)=127.934$, $\alpha_s(M_Z)=0.1172$ and $m_b(m_b)=4.25$~GeV
in the $\overline{MS}$ scheme, and an onshell top mass of $m_t=172.4$~GeV~\cite{Amsler:2008zzb}. 
Moreover, $M_Z=91.187$ and $m_\tau=1.777$~GeV, and $G_F=1.16639\cdot 10^{-5}$ GeV$^{-2}$.

To take into account the limits from direct SUSY searches at LEP \cite{lepsusy}, we require 
$m_{\tilde\chi^\pm_1}>103.5$~GeV and $m_{\tilde e,\tilde\mu}>100$~GeV. The limit on 
$m_{\tilde\tau_1}$ is parametrized as a function of $m_{\tilde\chi_1}$ as given by 
\cite{lepsusy}; in case of a stau LSP, we take $m_{\tilde\tau_1}>94$~GeV.
For the light scalar Higgs, we apply the limits from LEP for the $m_h^{max}$ scenario 
given in \cite{lephiggs}, taking into account a $\sim2$~GeV theoretical error \cite{Degrassi:2002fi}. 
\footnote{Moreover,   
the limits from direct squark and gluino searches at the Tevatron for 
$m_{\tilde q}\simeq m_{\tilde g}$ apply, in particular $m_{\tilde g}>392$~GeV,  
but these are automatically fulfilled here.}

We also take into account additional constraints from $B$-physics. 
For the branching ratio of inclusive radiative $B$ decay, we use the 
experimental result BR$(b\to s\gamma )=(3.52\pm 0.23\pm0.09)\times 10^{-4}$ from 
HFAG \cite{Barberio:2008fa}, 
together with the SM theoretical prediction of 
BR$(b\to s\gamma )^{\rm SM}=(3.15\pm 0.23)\times 10^{-4}$ of~\cite{Misiak:2006zs}. 
Combining experimental and theoretical errors in quadrature, we require 
$2.85\le {\rm BR}(b\to s\gamma)\times 10^4 \le 4.19$ at $2\sigma$.
Another important constraint comes from the $B_s$ decay into a pair of muons.
Here we apply the 95\% CL upper limit BR$(B_s\to\mu^+\mu^-)<5.8\times 10^{-8}$ from 
CDF~\cite{Aaltonen:2007kv}.  
Regarding the anomalous magnetic moment of the muon, we do not impose any limits 
but simply note that $(g-2)_\mu$ favours $\mu>0$.  

Last but not least, if the lightest neutralino is the LSP, we compare its relic density to 
the results from the 5-year WMAP data on the dark matter relic density, 
$\Omega h^2=0.1099\pm0.0062$~\cite{Dunkley:2008ie}, although we do not impose 
this as a strict constraint.  
The values of BR$(b\to s\gamma )$,  BR$(B_s\to\mu^+\mu^-)$ and $\Omega h^2$ 
are computed using the {\tt micrOMEGAs2.2} package \cite{Belanger:2006is}.

\section{Results for simplified boundary conditions}\label{sec:simplenum}

Here we perform a first exploration of the parameter space using simplified 
boundary conditions in the matter sector according to \cite{Choi:2003kq}. 
More precisely, we assume that not only the first two generations but also 
the third-generation leptons and r.h. bottom are brane-localized. The top
and l.h. bottom have a flat profile in the fifth dimension. The relevant 
kinetic functions then are\footnote{In 
this approximation $y_t=g_4$ is the only non-vanishing Yukawa coupling. We 
will however not enforce this in the numerical analysis.
}
\be\label{ys}
Y_{Q_3}\approx Y_{U_3}\approx\frac{\pi}{2}(T+\overline{T}),
\ee
which leads to
\be\label{rel}
m_{Q_3}^2\approx m_{U_3}^2\approx\left|\frac{F^T}{2R}\right|^2
\ee
and
\be\label{at}
A_t\approx\frac{F^T}{2R}\frac{1}{1+c'}.
\ee

\noindent
The setup for the first parameter scan is therefore as follows:
\begin{itemize}
\item We vary $M_{1/2}$ from 100 and 1000 GeV and $\tan\beta$ from 2 and 20. 
(For higher values of $\tan\beta$, the bottom Yukawa coupling would be no longer negligible.)
\item We set $m_{U_3}^2=m_{Q_3}^2=M_{1/2}^2(1+c')^2$ and $A_t=M_{1/2}$; this requires 
the additional iteration on $c'$ as detailed in Section \ref{sec:generalnum}.
All other sfermion soft terms are assumed to be zero at the GUT scale.
\item We allow for all four sign combinations of sign$(\mu)=\pm1$ and $\epsilon_{H}=\pm1$.
\item $\mu$ and $B\mu$ are determined from Eq.~\eqref{musugra} at the EWSB scale, while 
         $m_{H_1}^2$ and $m_{H_2}^2$ are determined from Eq.~\eqref{ghubound} at 
         $M_{\rm GUT}$.
\item For each point that gives correct EWSB, we check the mass limits from LEP as well as 
         the constraints from BR$(b\to s\gamma )$ and BR$(B_s\to\mu^+\mu^-)$ given in
	Section \ref{sec:generalnum}
\end{itemize}

\begin{figure}[t]\centering
   \includegraphics[width=8cm]{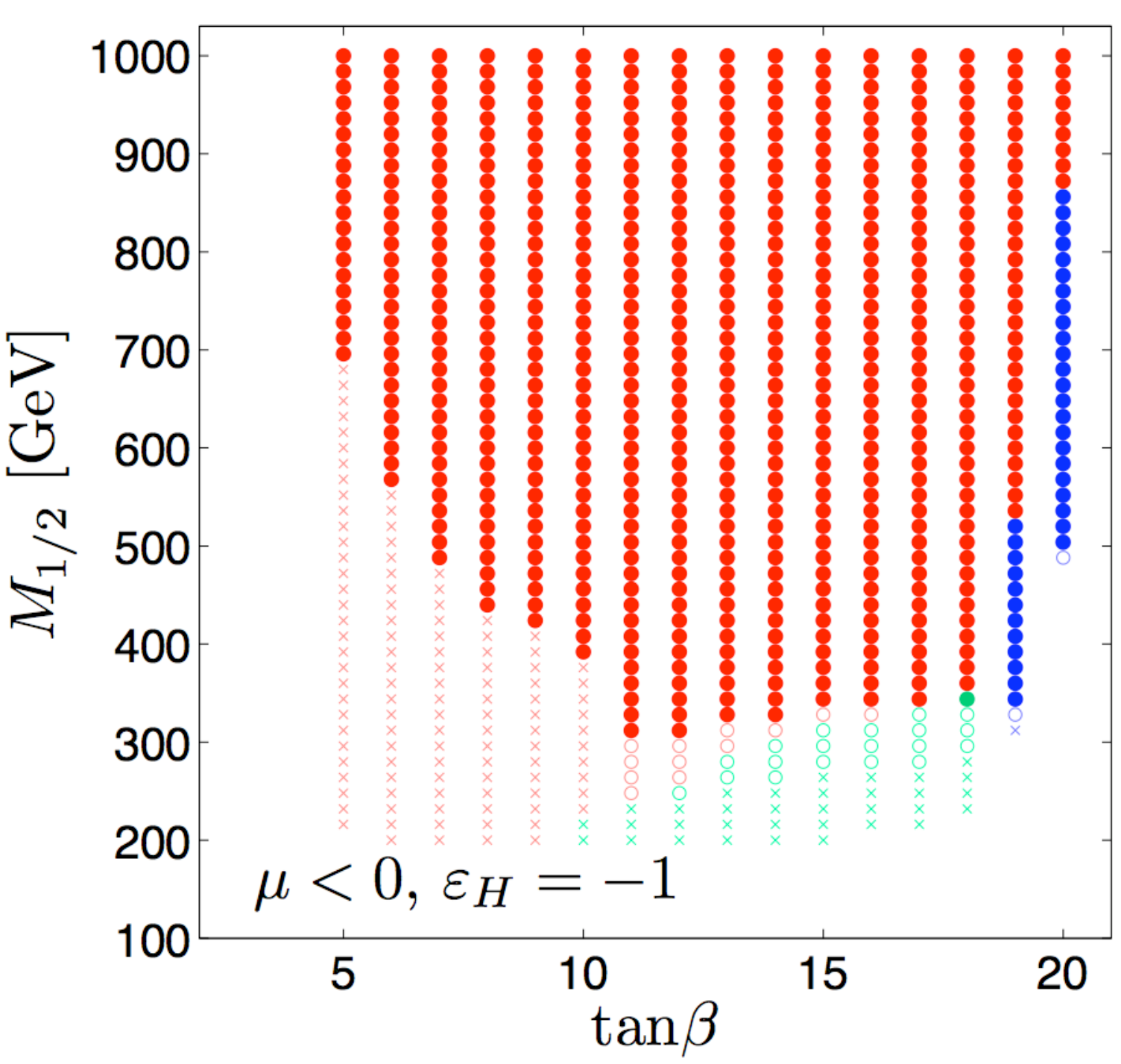} 
   \includegraphics[width=8cm]{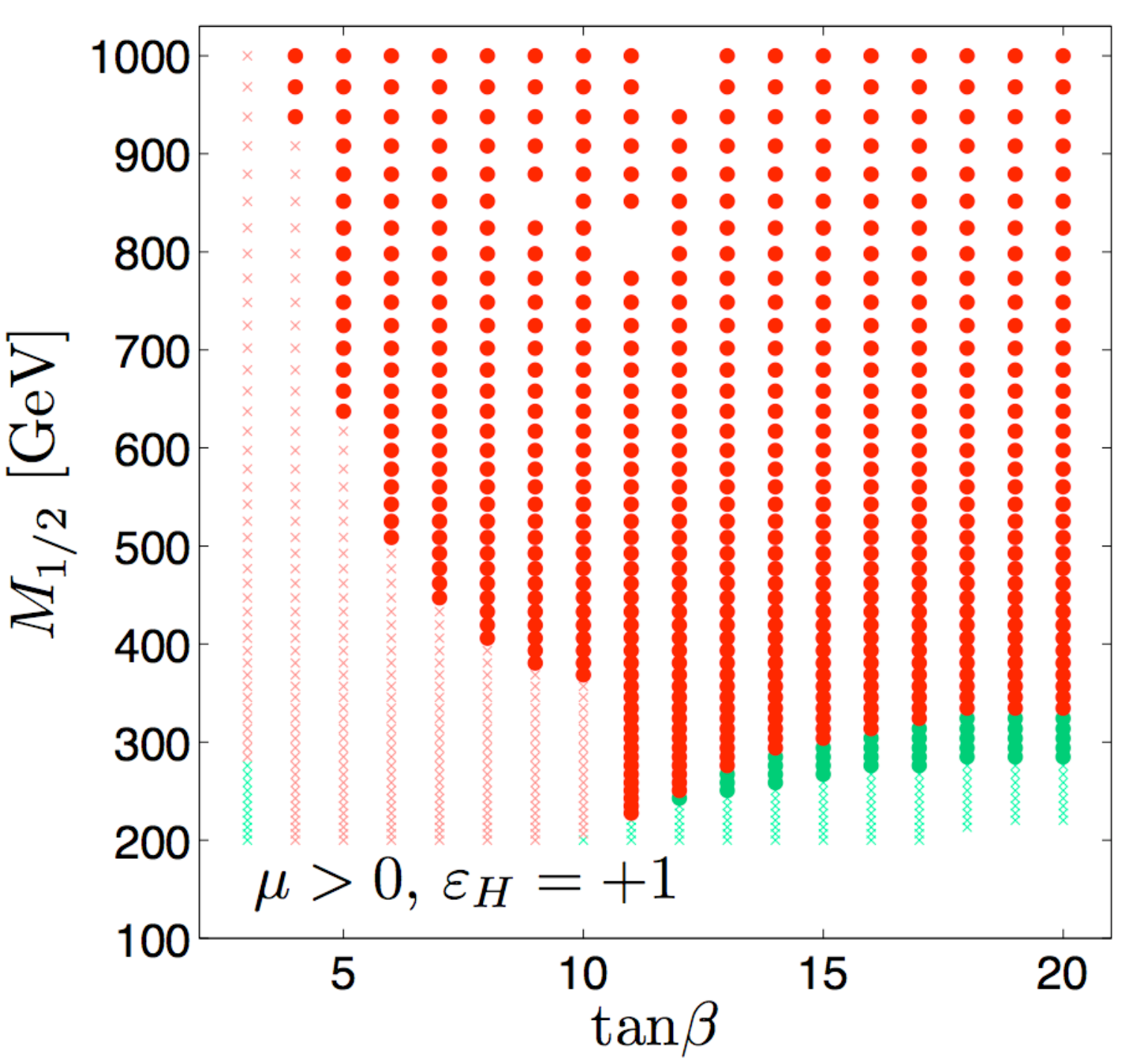} 
   \caption{Parameter points giving correct EWSB from a scan over $M_{1/2}$ 
   and $\tan\beta$ with simplified boundary conditions. 
   The red, green and blue points have 
   a neutralino, stau and selectron LSP, respectively. 
   Small crosses denote points excluded by LEP, while 
   open circles denote points excluded by B-physics constraints. 
   The big full dots pass these constraints.}
   \label{fig:BNsimp-tb-mhf}
\end{figure}

\noindent
Figure~\ref{fig:BNsimp-tb-mhf} shows the result of this scan in the $\tan\beta$ versus 
$M_{1/2}$ plane. 
As expected, correct EWSB is obtained only for two of the four possible combinations of 
sign$(\mu)$ and $\epsilon_{H}^{}$. In particular, it turns out that the two signs need 
to be equal. This is a consequence of the relation $A_t=M_{1/2}$, in accord with the 
discussion in Section~\ref{sec:running}. 
Phenomenological bounds further constrain the parameter space. 
Points marked as small crosses in Fig.~\ref{fig:BNsimp-tb-mhf} are excluded by the 
mass bounds from LEP, while points shown as open circles are excluded by 
BR$(b\to s\gamma )$; the constraint from BR$(B_s\to\mu^+\mu^-)$ has no effect. 
The remaining big full points are phenomenologically viable.
The different colours denote the nature of the LSP: red for a neutralino, blue for a selectron
\footnote{Selectrons 
and smuons are taken to be mass-degenerate. Here and in the following we only refer to 
selectrons for simplicity, implicitly meaning ``selectrons and smuons''.}, 
and green for a stau LSP. As one can see, most of the parameter space features a neutralino 
LSP, which is interesting in point of view of dark matter.\footnote{Alternative dark matter 
candidates would be gravitino or axino. 
A rough estimate for a no-scale radion K\"ahler potential $K=-3\log(T+\ol T)$ 
gives $m_{3/2}>|F^T/2R|$, while $m_{\tilde\chi^0_1}\simeq 0.4\,M_{1/2}$. 
In this case a gravitino LSP is only possible for $c'<-0.6$, which does not occur in our analysis. 
Moreover, we expect other contributions from hidden sectors to further increase $m_{3/2}$.  
An axino LSP is a valid option, but leads to a very different phenomenology, beyond 
the scope of this paper.} 
As anticipated in Section \ref{sec:running}, $|\mu|$ turns out to be large throughout the 
parameter space. Numerically we find 
$|\mu|\sim (2.5-3.5)M_{1/2}$ for $\mu>0$ and $|\mu|\sim (2.5-4)M_{1/2}$ for $\mu<0$; 
in both cases the values at the high end are obtained for larger $\tan\beta$.
The $\tilde\chi^0_1$ is hence almost a pure bino, and the  $\tilde\chi^0_2$ and 
$\tilde\chi^\pm_1$ almost pure winos.

\begin{figure}[tp]\hspace*{-4mm}\begin{tabular}{rr}
   \includegraphics[width=7.9cm]{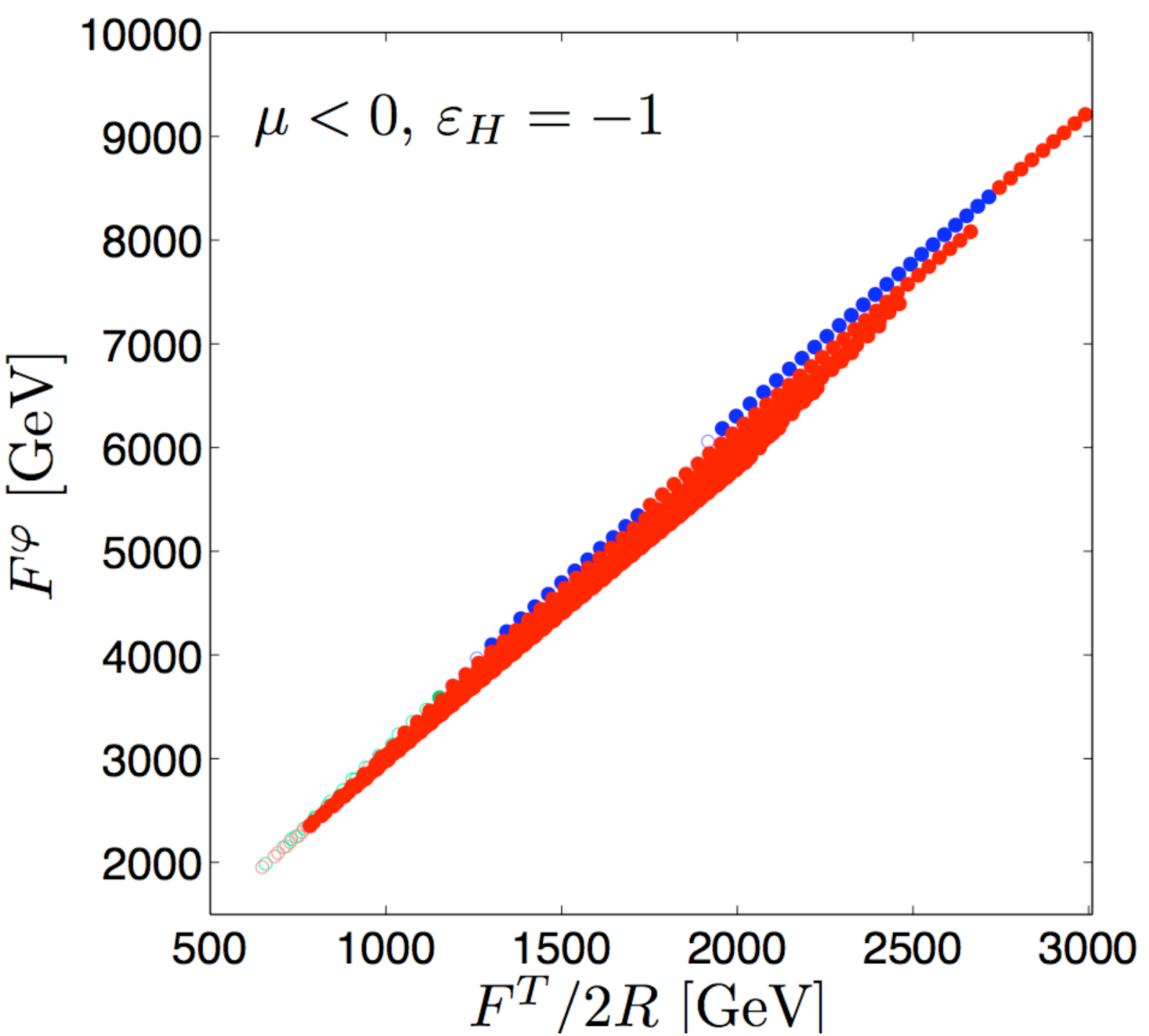}& 
   \includegraphics[width=7.9cm]{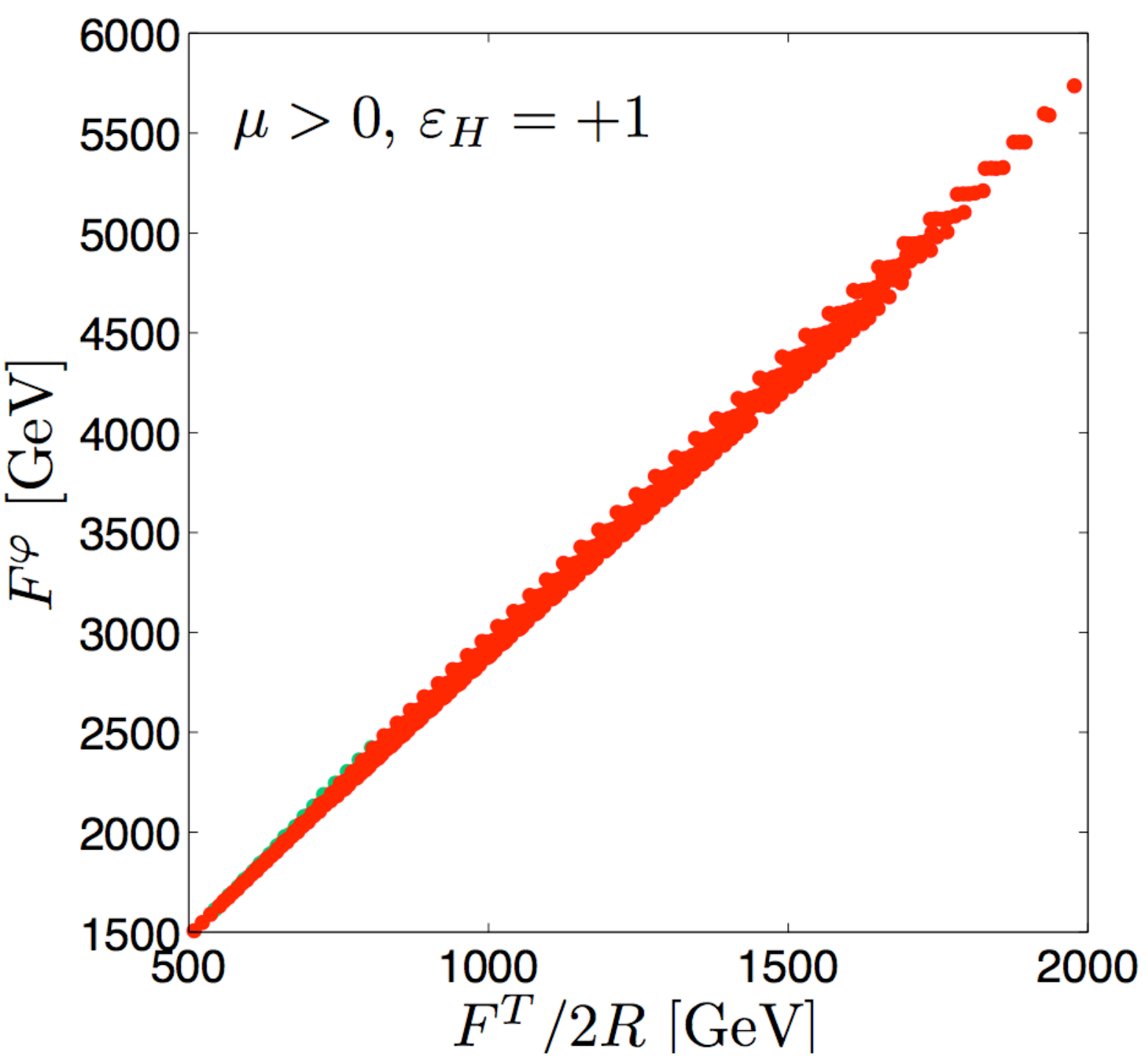}\\
   \includegraphics[width=7.7cm]{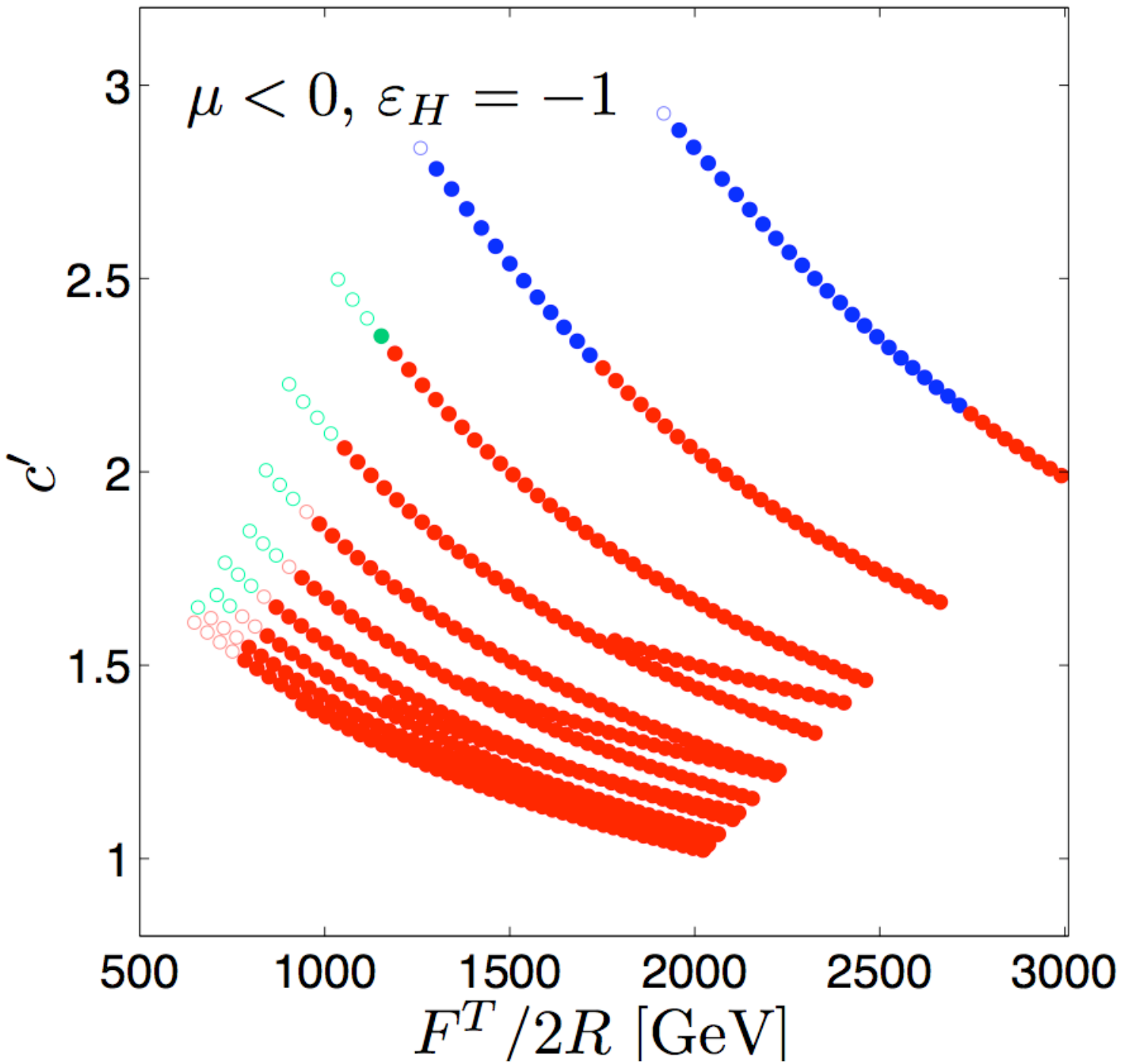}&
   \includegraphics[width=7.7cm]{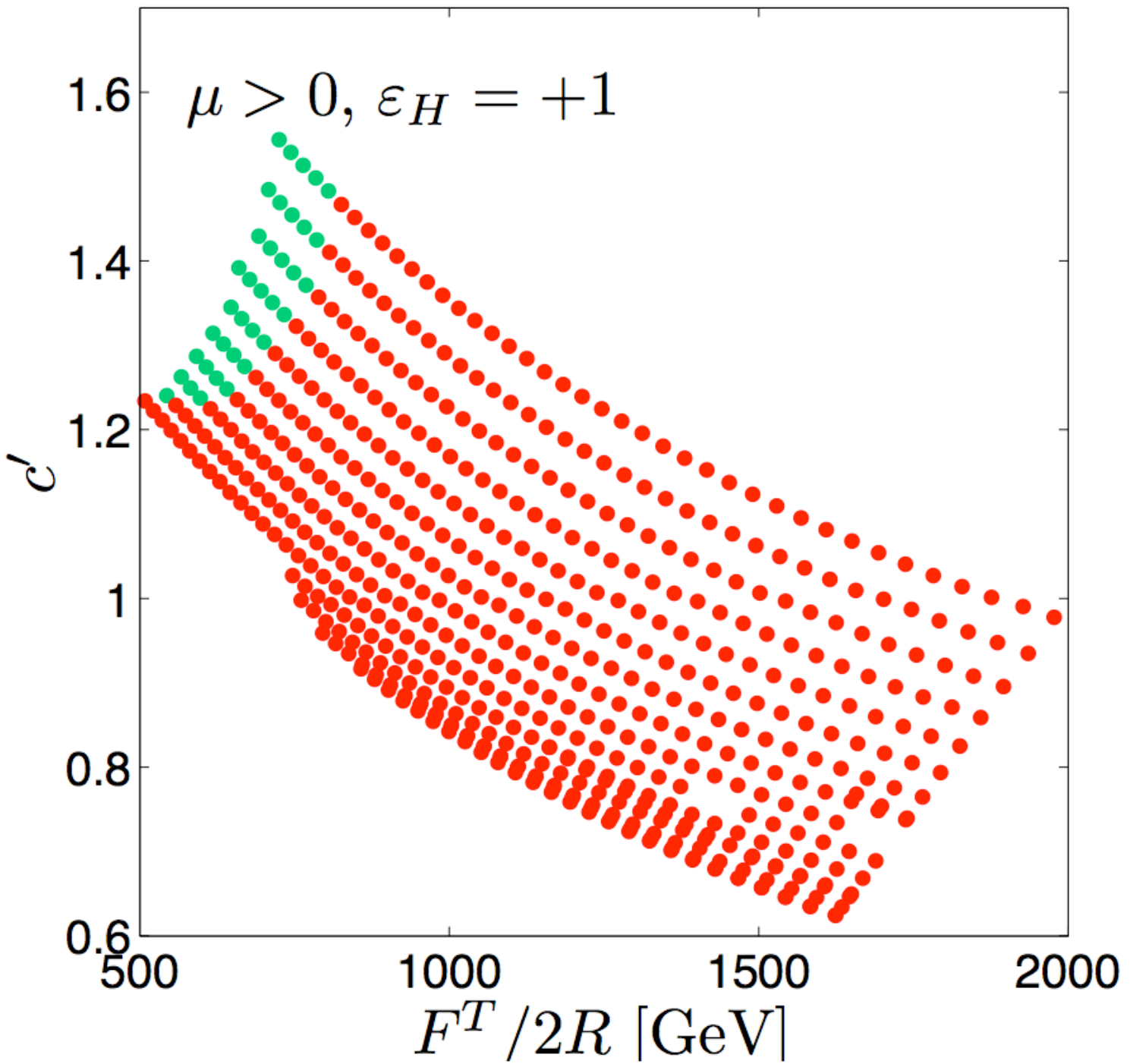}
   \end{tabular}
   \caption{Scatter plot of points which give a valid spectrum solution 
   in the $F^T/2R$ vs.\ $F^\varphi$
   plane (top row) and in the $F^T/2R$ vs.\ $c'$ plane (bottom row). 
   The red, green and blue points have a neutralino, stau and selectron LSP, respectively.
   Open circles denote points excluded by B-physics constraints. 
   Points excluded by LEP are not shown.}
   \label{fig:BNsimp-parspace}
\end{figure}

The projections onto the space of fundamental model parameters 
$F^T/2R$, $F^\varphi$ and $c'$ are shown in Fig.~\ref{fig:BNsimp-parspace}. 
We observe that for both, $\mu<0$ and $\mu>0$, there is a strong correlation between 
$F^\varphi$ and $F^T/2R$, with roughly $F^\varphi \sim 3\times F^T/2R$. 
This comes from setting $A_t=M_{1/2}$, which enforces $\epsilon_H^{}={\rm sign}(\mu)$.
It translates into a large value of $F^\varphi$, because 
$F^\varphi=\epsilon_H^{}\mu+ F^{T}/2R\,\frac{1+2c'}{1+c'}$ from Eq.~\eqref{mubmu1}. 
Nevertheless  $F^\varphi$ is small enough so that contributions from anomaly mediation, 
being ${\cal O}(F^\varphi/8\pi^2)$, can safely be neglected. 

It is particularly interesting to note that we find no valid spectra for which $c'=0$. 
This also holds when considering points excluded by LEP constraints.
In this sense our analysis confirms the result of \cite{Choi:2003kq}, who did not include
the effects of a Chern--Simons term and consequently did not find any viable 
parameter regions, except for extremely unnatural values for the gaugino masses 
far above our scan limits. At the same time it is important that $c'$, which is an  
${\cal O}(1)$ parameter, never becomes large. 

\begin{figure}[t]\centering
   \includegraphics[width=8cm]{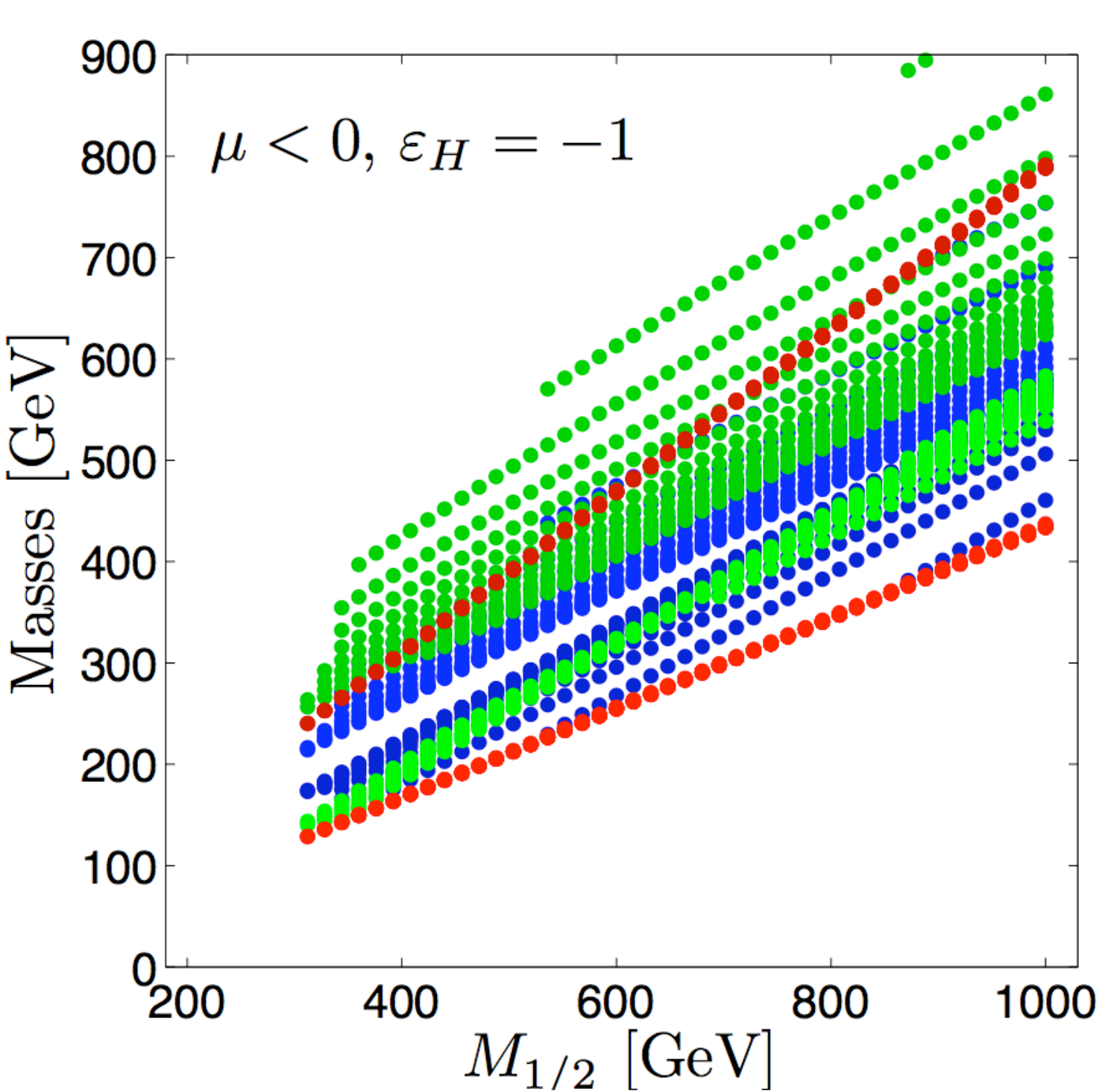} 
   \includegraphics[width=8cm]{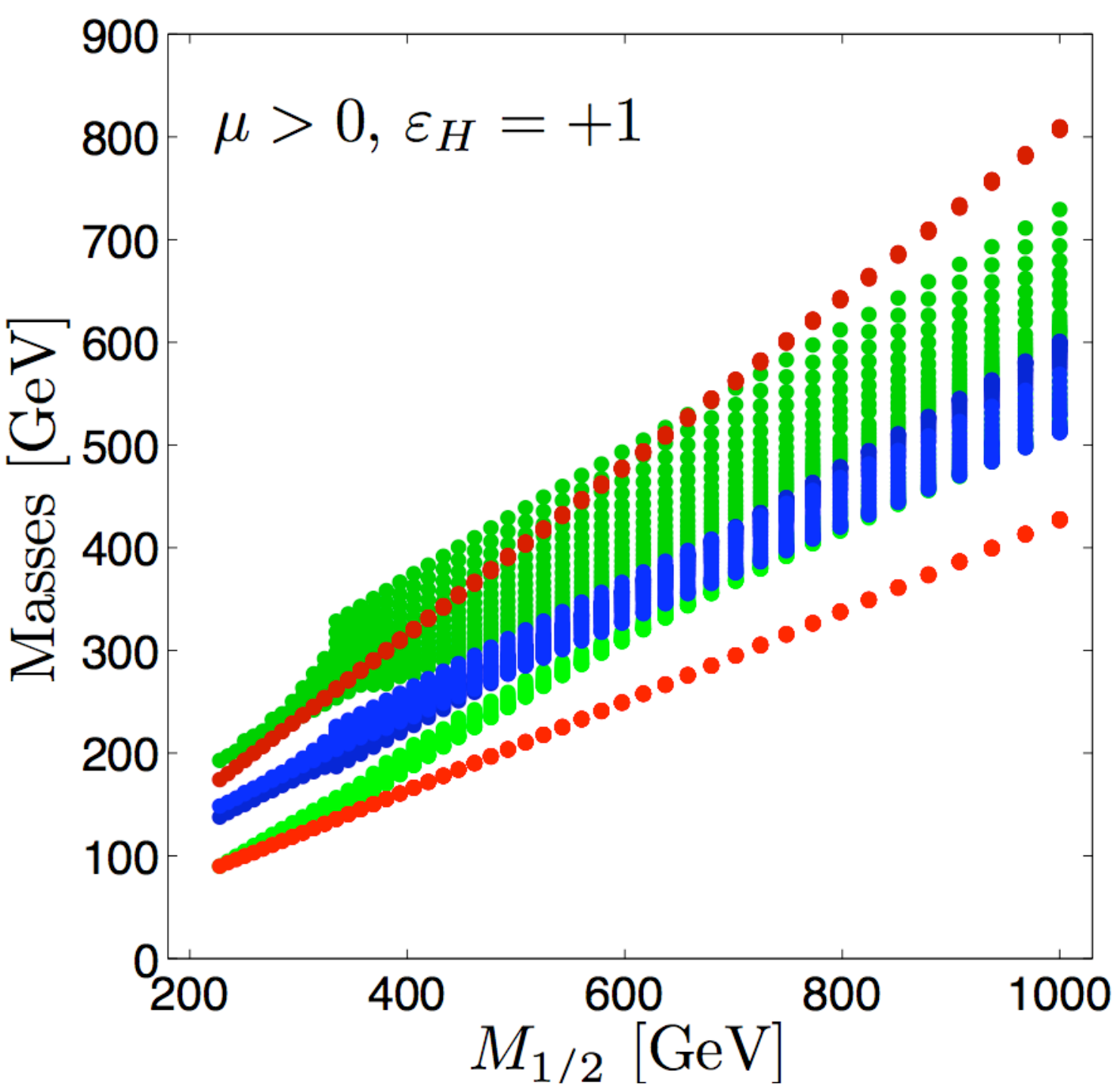} 
   \caption{Mass spectrum in the neutralino LSP region, passing LEP and B-physics constraints, 
   as a function of $M_{1/2}$.    The colour convention is as follows: 
   red: $\tilde\chi^0_1$, green: $\tilde\tau_1$, blue: $\tilde e_R$, dark blue: $\tilde e_L$, , 
   dark green: $\tilde\tau_2$, dark red: $\tilde\chi^0_2$.}
   \label{fig:BNsimp-masses}
\end{figure}

Implications for collider phenomenology can be deduced from Fig.~\ref{fig:BNsimp-masses}, 
which shows the neutralino and slepton mass spectrum in the neutralino LSP region. 
We see that the second-lightest neutralino $\tilde \chi^0_2$ and the lighter chargino 
$\tilde\chi^\pm_1$, which are mainly winos ($m_{\tilde\chi^\pm_1}\simeq m_{\tilde\chi^0_2}$), 
are always heavier than $\tilde e_{R,L}$ and $\tilde\tau_1$ 
(with the exception of a few points at $\mu<0$ which have 
$m_{\tilde e_R}>m_{\tilde\chi^0_2}>m_{\tilde e_L}$).  
Note the clear separation of the selectron masses with $m_{\tilde e_L}<m_{\tilde e_R}$ for 
$\mu<0$, while for $\mu>0$ we have $m_{\tilde e_L}\sim m_{\tilde e_R}$.
The squark and gluino masses are not shown, but they are roughly 
$m_{\tilde q}\sim m_{\tilde g}\sim(2-3)M_{1/2}$. 
At the LHC, squarks and gluinos will hence be produced 
both as $\tilde q\tilde q$ or $\tilde g\tilde g$ pairs,  
and in $\tilde q\tilde g$ associated production. Their decays are $\tilde g\to q\tilde q_{R,L}^{}$, 
$\tilde q_R^{} \to q\tilde\chi^0_1$, $\tilde q_L^{}\to q'\tilde\chi^\pm_1$ or $q\tilde\chi^0_2$, 
as in the mSUGRA scheme with large $|\mu|$~\cite{Baer:1995nq}. 
Moreover, the decays  
$\tilde\chi^0_2 \to e^\pm \tilde e_L^\mp \to e^+ e^- \tilde\chi^0_1$ and 
$\tilde\chi^0_2 \to \tau^\pm \tilde\tau_1^\mp \to \tau^+ \tau^- \tilde\chi^0_1$ 
are always open and together have about 50\% branching ratio; the other 50\% go into 
neutrinos. This leads to the gold-plated same-flavour opposite-sign (SFOS) dilepton signature 
at the LHC~\cite{Baer:1995va}, which allows to reconstruct sparticle masses.
Furthermore, the decay of the lighter chargino always leads to a charged lepton, 
$\tilde\chi^\pm_1\to (\ell^\pm \tilde\nu_{\ell}\ {\rm or}\ \nu_\ell \tilde \ell^\pm_L) 
\to \ell^\pm \nu_\ell \tilde\chi^0_1$, 
giving rise to a large number of 
events with jets plus 1 hard lepton plus missing transverse energy, $E_T^{\rm miss}$. 
If combined with $\tilde\chi^0_2\to ... \to l^+l^-\tilde\chi^0_1$, this leads to the rather clean 
trilepton signature (plus jets plus $E_T^{\rm miss}$). 

The scenario becomes even more predictive if we require that the neutralino LSP have 
a relic density in agreement with cosmological observations (assuming standard cosmology).
Imposing the $3\sigma$ upper bound from WMAP5, $\Omega h^2<0.1285$, 
constrains $M_{1/2}\lesssim 390$~GeV with $\tan\beta\gtrsim11$ for $\mu>0$. 
For $\mu<0$, it gives an upper limit on $M_{1/2}$ which increases with $\tan\beta$, 
from $M_{1/2}\lesssim 312$~GeV at $\tan\beta=12$ to  
$M_{1/2}\lesssim 920$~GeV at $\tan\beta=20$.
The reason is that the LSP is almost a pure bino and has a small pair-annihilation 
cross section (s-channel Higgs exchange is not efficient in this scenario); 
in order to have a small enough relic density, the LSP needs to co-annihilate with another 
sparticle which is close in mass, typically the next-to-lightest SUSY particle (NLSP).  
This constrains the scenario to the region of small NLSP--LSP mass differences
near the boundary to the slepton LSP region, which is realized  
for $\mu<0$ up to large $M_{1/2}$ (depending on $\tan\beta$), 
but for $\mu>0$ only at small $M_{1/2}$, cf.~Figs.~\ref{fig:BNsimp-tb-mhf} and 
\ref{fig:BNsimp-masses}. Note, however, that this is a direct consequence of the 
simplified assumptions for the matter sector.

\section{Realistic sfermion soft terms}\label{sec:realistic}

In this section we explain how improved sfermion soft terms can be
obtained if we model the matter sector as in the Burdman--Nomura 
model~\cite{Burdman:2002se}.
The third generation matter fields arise from the mixing of brane and bulk 
fields. Bulk fields with flat profile have Yukawa couplings determined by the 
5d gauge coupling. Non-trivial bulk profiles cause a reduced overlap with 
the Higgs wave function and hence smaller Yukawa couplings. Thus, using 
both bulk masses and mixing angles we can obtain realistic values for $y_t$, 
$y_b$ and $y_\tau$.

For the third generation quarks in particular, we introduce a 5d 
bulk hypermultiplet $\{{\cal U},{\cal U}^c\}$ in the ${\bf 20}$ of SU$(6)$ 
containing as 4d zero modes the right-handed top quark superfield and a weak 
doublet, and another bulk hypermultiplet $\{{\cal D},{\cal D}^c\}$ in the 
${\bf 15}$ containing the right-handed bottom quark and a second doublet. We 
give these fields bulk masses $M_u$ and $M_d$. Furthermore, brane-localized 
superfields must be introduced to decouple unwanted massless fields. They 
couple to the doublet components of both the ${\cal U}$ and ${\cal D}$ fields, 
leaving a single massless quark doublet instead of the two we were starting with.
 This effect is parametrized by a mixing angle $\phi_Q$. 

Similarly, leptons descend from two 5d bulk hypermultiplets, 
$\{{\cal E},{\cal E}^c\}$ in the ${\bf 15}$ and
$\{{\cal N},{\cal N}^c\}$ in the ${\bf 6}$. In analogy with the quark sector this 
leads to three more model parameters, two bulk masses $M_e$ and $M_n$ and a mixing 
angle $\phi_L$. For details of the model, in particular for the proper choice of 
boundary conditions, brane fields and bulk-brane couplings, we refer to 
\cite{Burdman:2002se}.

The kinetic functions are computed by integrating the zero-mode profiles over 
the fifth dimension, replacing its radius $R$ by $(T+\overline{T})/2$. This gives
\begin{align}
\label{YU}Y_{U_3}&=\frac{1}{2|M_u|}\left(1-e^{-\pi(T+\overline{T})|M_u|}\right),\\
\label{YQ}Y_{Q_3}&=\frac{1}{2|M_u|}\left(1-e^{-\pi(T+\overline{T})|M_u|}\right)
\sin^2(\phi_Q)+\frac{1}{2|M_d|}\left(1-e^{-\pi(T+\overline{T})|M_d|}\right)\cos^2(\phi_Q),\\
\label{YD}Y_{D_3}&=\frac{1}{2|M_d|}\left(1-e^{-\pi(T+\overline{T})|M_d|}\right).
\end{align}
The kinetic functions for the lepton sector are obtained in the same manner, and
are given by the same expressions with the obvious parameter replacements.
The soft masses and $A$-terms are then derived from Eqs.~\eqref{eq:mx}
-- \eqref{eq:ae}. We refrain from giving 
closed-form expressions for them, since these are rather cumbersome and not very 
illuminating.

The parameters $M_u, M_d, M_n, M_e, \phi_Q$ and $\phi_L$ cannot be chosen entirely
freely, because they also have to account for the proper physical values of
the Yukawa and gauge couplings. Since the Higgs wave function normalization is 
just given by $\langle Y_H\rangle=1/g_4^2$ and in particular is independent of $c'$, 
the relations given in \cite{Burdman:2002se} apply:\footnote{Note that our conventions 
for $\phi_Q$ and $\phi_L$ slightly differ from those of \cite{Burdman:2002se}.}
\be\label{yukawasQ}
y_t = \sin(\phi_Q)\,\frac{\pi R|M_u|}{\sinh{\pi R|M_u|}}\,g_4,, \quad 
y_b = \cos(\phi_Q)\,\frac{\pi R|M_d|}{\sinh{\pi R|M_d|}}\,g_4\,,
\ee
\be\label{yukawasL}
y_n = \sin(\phi_L)\,\frac{\pi R|M_n|}{\sinh{\pi R|M_n|}}\,g_4\,, \quad
y_\tau = \cos(\phi_L)\,\frac{\pi R|M_e|}{\sinh{\pi R|M_e|}}\,g_4\,.
\ee

In the numerical analysis, in order to avoid additional model dependence from 
the unknown neutrino sector, we will assume that $M_n$ is large enough not 
to contribute to the stau soft terms.  We also introduce a Majorana
mass term for the right-handed neutrinos on the $y=0$ brane as in 
\cite{Burdman:2002se}. Since $M_n$ is large, the neutrino wave function will be
strongly localized towards the $y=\pi R$ brane, resulting in an exponentially
suppressed Yukawa coupling and a doubly exponentially suppressed Majorana
mass. The suppression factors will cancel out in the see-saw formula for the
lighter neutrino mass eigenstate, leading to the same lighter neutrino mass as 
in the standard see-saw mechanism. The heavier neutrino mass, on the other
hand, will be lowered by a factor $\sim e^{-4\pi R |M_n|}$ with respect
to the GUT scale. This may be beneficial for leptogenesis 
\cite{Hebecker:2002xw}.

It is instructive to see how Eq.~\eqref{yukawasQ} constrains the possible
ranges of squark soft terms. In the remainder of this section we
will therefore give some estimates of the bounds on the squark masses and
trilinear couplings.

For $\tan\beta\sim 5-50$, the relevant GUT-scale Yukawa couplings 
take values $0.5\lesssim y_t\lesssim 0.6$ and $0.02\lesssim y_b\lesssim 0.3$. 
We also know that the gauge couplings unify at $g_4\approx 0.7$. 

\begin{figure}[t]\centering
   \includegraphics[width=8cm]{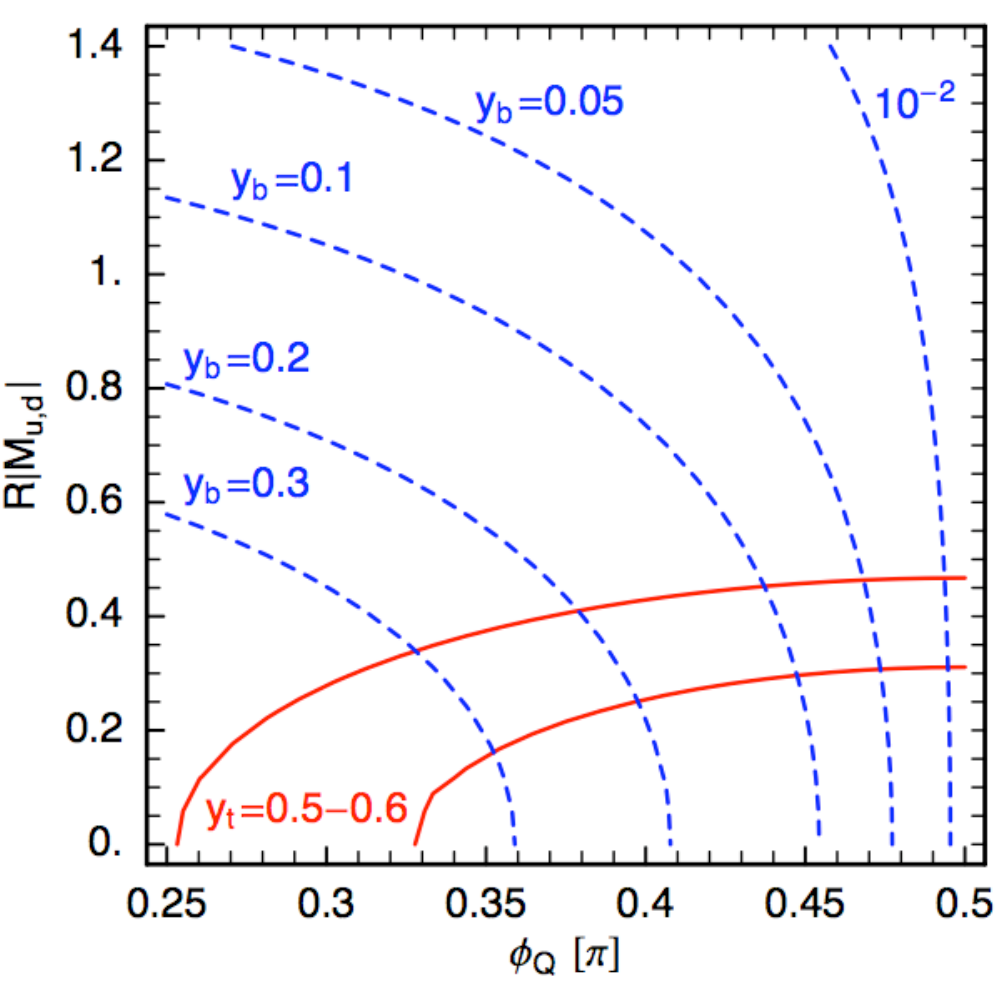} 
   \caption{Values of $R|M_u|$ and $R|M_d|$ as function of the mixing angle $\phi_Q$ 
   for various values of $y_t$ (full red lines) and $y_b$ (dashed blue lines). Note that 
   $y_t$ gives the lower and $y_b$ the upper bound of the allowed range of  $\phi_Q$.}
   \label{fig:BulkMass}
\end{figure}

To reproduce the top Yukawa coupling, we must have $\tan\phi_Q\gtrsim 1$ by 
Eq.~\eqref{yukawasQ}. The small ratio $y_b/y_t$ can then be generated either 
by choosing $\tan\phi_Q$ to be large,  or choosing $|M_d| > |M_u|$, or by a 
combination of these. The relation between $M_{u}$, $M_{d}$, $\phi_Q$ and the 
Yukawa couplings is illustrated in Fig.~\ref{fig:BulkMass}. We note that for 
given $y_t$ and $y_b$, the allowed range for the mixing angle $\phi_Q$ is
\be
  \phi_Q =\big[ \arcsin(y_t/g_4),\:\arccos(y_b/g_4)\, \big] .
\ee

\noindent
For estimating the size of the squark-mass parameters, let us consider two 
limiting cases:
\begin{itemize}
 \item If the difference between $y_t$ and $y_b$ is mainly due to the 
different bulk masses, then $\tan\phi_Q\approx 1$. This corresponds to the 
far left region of Fig.~\ref{fig:BulkMass}. In that case 
$\sin\Phi_Q\approx 1/\sqrt{2}$ already accounts for the ratio 
$y_t/g_4\approx 0.7$ in Eq.~\eqref{yukawasQ}. The top Yukawa coupling should 
thus not receive much additional suppression from large bulk masses, hence 
we need $|M_u|\ll 1/R$. Expanding Eq.~\eqref{YU} and retaining only the 
leading term, we reproduce $Y_{U_3}$ as in Eq.~\eqref{ys}:
\be
Y_{U_3}=\frac{\pi}{2}\left(T+\ol T\right),\qquad m_{U_3}^2
=\left|\frac{F^T}{2R}\right|^2.
\ee

On the other hand, $R |M_d|$ must be sizeable to obtain an appropriately 
suppressed $y_b$, cf.\ Fig.~\ref{fig:BulkMass}. With Eq.~\eqref{YD}, 
$m_{D_3}^2$ turns out to be 
\be
m_{D_3}^2=\left(\frac{\pi R |M_d|}{\sinh\left(\pi R |M_d|\right)}\right)^2
\left|\frac{F^T}{2R}\right|^2\approx 4y_b^2\left|\frac{F^T}{2R}\right|^2.
\ee

Finally, the quark doublet soft mass-squared $m_{Q_3}^2$ obtained from 
Eq.~\eqref{YQ} is numerically
\be
m_{Q_3}^2\approx (0.7-0.8)\times \left|\frac{F^T}{2R}\right|^2.
\ee

\item If $\tan\phi_Q\gg 1$, i.e.~$\sin\phi_Q\approx 1$ (which is the case
in the far right region of Fig.~\ref{fig:BulkMass}), then the ratio 
$y_t/g_4\approx 0.7$ is mainly due to a sizeable bulk mass $M_u$. Numerically, 
we need $ R|M_u|\approx 0.3-0.5$. Therefore we should use the full 
expression for $Y_{U_3}$, rather than just the leading term:
\be
m_{U_3}^2=\left(\frac{\pi R |M_u|}{\sinh\left(\pi R|M_u|\right)}\right)^2\left|\frac{F^T}{2R}\right|^2\approx (0.5-0.8)\times\left|\frac{F^T}{2R}\right|^2.
\ee
Dropping the $\cos^2\phi_Q$ piece in $Y_{Q_3}$ and setting $\sin\phi_Q=1$, 
we obtain the same expression for $Y_{Q_3}$ and eventually $m_{Q_3}^2$:
\be
m_{Q_3}^2=\left(\frac{\pi R |M_u|}{\sinh\left(\pi R|M_u|\right)}\right)^2\left|\frac{F^T}{2R}\right|^2\approx (0.5-0.8)\times\left|\frac{F^T}{2R}\right|^2.
\ee
As is evident from Fig.~\ref{fig:BulkMass}, if $y_b$ is to remain finite, 
$\tan\phi_Q$ cannot become arbitrarily large. In any case, this limit requires
very small $y_b$. The constraints on $|M_d| R$ are rather weak, although
smaller $|M_d| R$ is somewhat favoured in order not to get additional $y_b$
suppression. Hence
\be
m_{D_3}^2\lesssim\left|\frac{F^T}{2R}\right|^2.
\ee 
\end{itemize}

\noindent 
In the end we expect the squark masses-squared to lie somewhere in between 
these two extremes:
\be
0.5\times\, \left|\frac{F^T}{2R}\right|^2\lesssim\left(m_{Q_3}^2,\, m_{U_3}^2\right)\lesssim  \left|\frac{F^T}{2R}\right|^2,\quad 0\lesssim m_{D_3}^2\lesssim \left|\frac{F^T}{2R}\right|^2.
\ee

\noindent
In order to obtain limits on $A_t$, we can make the same case distinction:
\begin{itemize}
 \item for $\tan\phi_Q\approx 1$ and small $|M_u|$, we get
\be
  A_t\approx \frac{F^T}{2R}\left(-\frac{1+2c'}{1+c'} + 1 
  + \frac{2\pi R|M_d|\,\left(1+e^{-2\pi R|M_d|}\right)}{2\pi R|M_d| +1-e^{-2\pi R|M_d|}}\right) ;
\ee
\item for $\sin\phi_Q\approx 1$, we obtain
\be
  A_t\approx \frac{F^T}{2R}\left(-\frac{1+2c'}{1+c'} + 
  2\frac{2\pi R |M_u|}{\exp\left(2\pi R|M_u| \right)-1}\right) .
\ee
\end{itemize}

Numerically,
\be
A_t\approx\frac{F^T}{2R}\left(-\frac{1+2c'}{1+c'}+\alpha \right)
\ee
where $0.3\lesssim\alpha\lesssim 2$, with $\alpha=2$ corresponding to the first of the above two cases (with $R|M_d|\approx 1$), and $\alpha = 0.3$ to the second (with $R|M_u| = 0.5$). Evidently $A_t$ can take a wide range of values, significantly departing from the simplified case of 
Section~\ref{sec:simplenum}. In particular it can become large and negative, which will be of 
relevance in the next Section.
A similar statement turns out to be true for $A_b$, for which we find an analogous estimate
with $0\lesssim\alpha\lesssim 1.4$.

\section{Results for realistic sfermion soft terms}\label{sec:bnnum}

Let us finally investigate to what extent the phenomenological features found 
in Section~\ref{sec:simplenum} remain valid when invoking realistic stop, 
sbottom and stau parameters derived from the Burdman--Nomura model. 
The six new parameters $M_u$, $M_d$, $M_n$, $M_e$, $\phi_{Q}$, $\phi_{L}$ 
are subject to four constraints, since 
they are related to the Yukawa couplings according to Eqs.~\eqref{yukawasQ} 
and \eqref{yukawasL}. 
As detailed above we assume that $M_n$ is large enough not to affect the stau 
soft terms. This corresponds to a negligible neutrino Yukawa coupling,  and 
we do not need to worry about lepton flavour violation~\cite{Borzumati:1986qx}. 
The precise value of $M_n$ is irrelevant.
(If $M_n$ did contribute to the stau soft terms, its main effect would be to 
increase $m_{L_3}$, thus rendering the staus heavier, but leaving the overall 
picture intact.) We are therefore left with five parameters, $M_{u,d,e}$ 
and $\phi_{Q,L}$, and three constraints from $y_t$, $y_b$ and $y_\tau$. 

We choose $\phi_Q$ and $\phi_L$ as the  two independent new parameters
and scan the parameter space as in Section~\ref{sec:simplenum}, with the 
following modifications:

\begin{itemize}

\item We vary $M_{1/2}$ from 100 to 1000~GeV, $\phi_Q$ from $\pi/4$ to $\pi/2$, 
and $\phi_L$ from $0$ to $\pi/2$. For $\tan\beta$, we consider three distinct 
values, $\tan\beta=10,\,20,$ and $30$, in order to avoid excessive computing 
times. 

\item For each point, 
the bulk masses $M_{u,d,e}$ are computed from the GUT-scale gauge and Yukawa 
couplings $g_4$, $y_t$, $y_b$, $y_\tau$ by numerically inverting 
Eqs.~\eqref{yukawasQ} and \eqref{yukawasL}. 
They then serve as input in the kinetic functions Eqs.~\eqref{YU}--\eqref{YD}, 
and the analogous expressions for the leptons, from which the sfermion soft 
masses and $A$-terms are obtained according to Eqs.~\eqref{eq:mx}--\eqref{eq:ae}.
The soft terms of the first and second generation are again assumed to be zero 
at the GUT scale. 

\end{itemize}

\begin{figure}\centering
   \includegraphics[width=15cm]{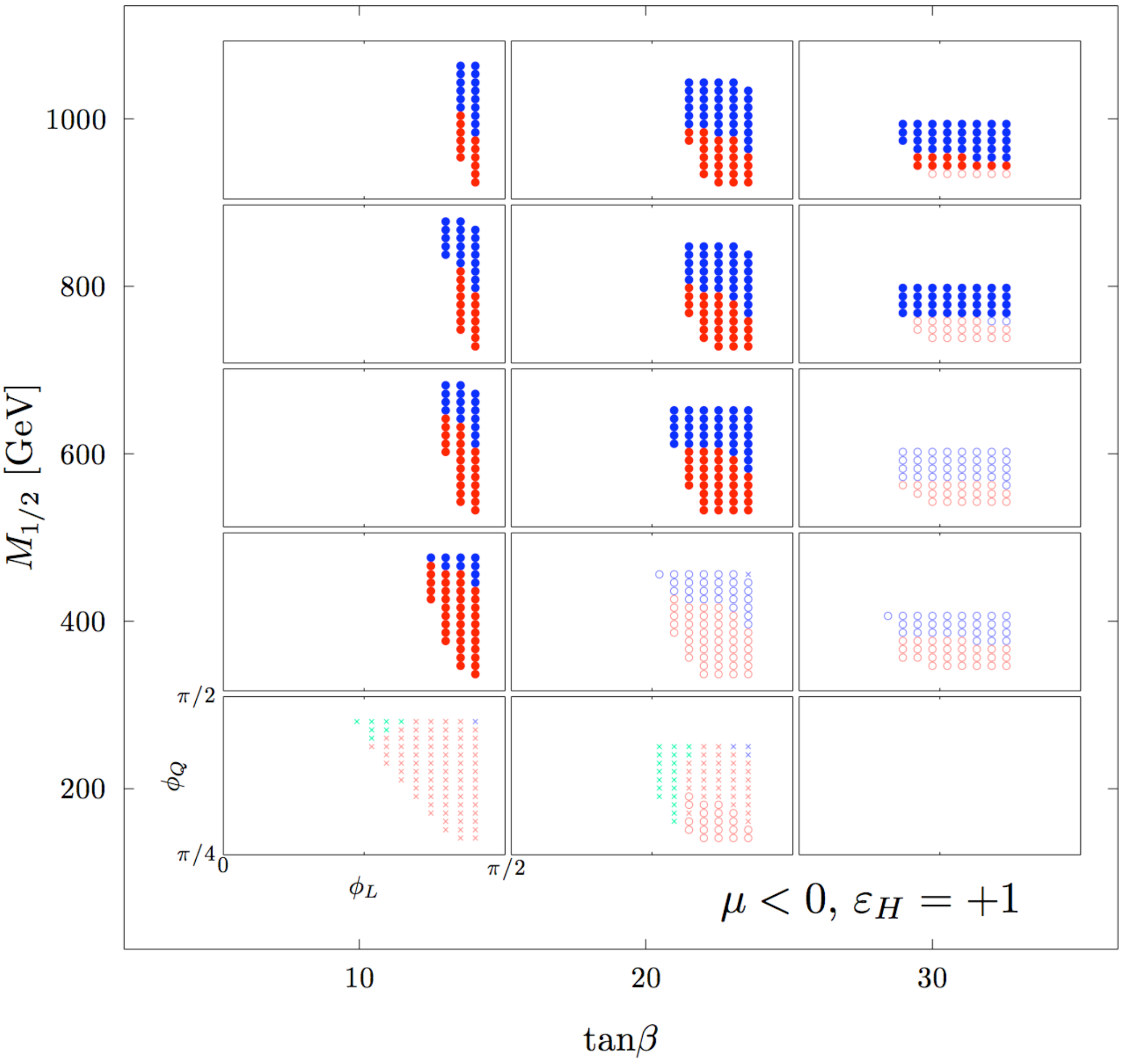}
   \caption{Points which lead to correct EWSB from a scan over $M_{1/2}$, $\tan\beta$, 
   $\phi_Q$ and $\phi_L$, for $\mu<0$, $\epsilon_{H}=+1$ and  
   sfermion soft terms determined according to the Burdman--Nomura model. 
   Small crosses denote points excluded by LEP, while 
   open circles denote points excluded by $B$-physics constraints. 
   Points passing these constraints are shown as big full dots.  
   The colours denote the nature of the LSP: red, green and blue points 
   have a neutralino, stau and selectron LSP, respectively.}
   \label{fig:BN-multiplot1}
\end{figure}

\begin{figure}\centering
   \includegraphics[width=15cm]{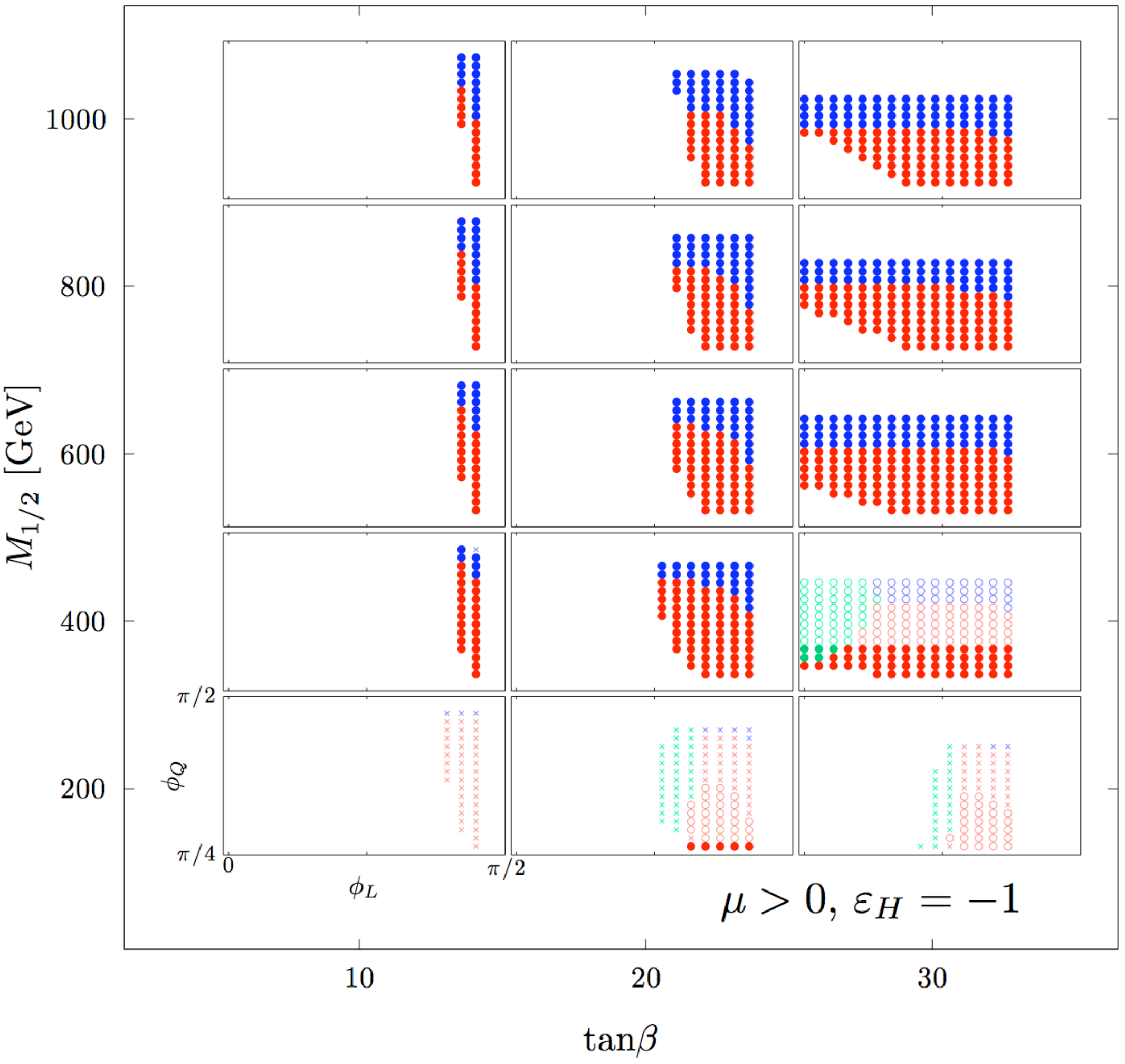}
   \caption{Same as Fig.~\ref{fig:BN-multiplot1} but for $\mu>0$ and $\epsilon_{H}=-1$.}
   \label{fig:BN-multiplot2}
\end{figure}

The result of this scan is shown in Figs.~\ref{fig:BN-multiplot1} and 
\ref{fig:BN-multiplot2} for the two signs of $\mu$. For better readability, we 
only  show $M_{1/2}$ in steps of 200 GeV, although the scan had a much finer 
grid. Contrary to the case of simplified boundary conditions, now $\mu$ and 
$\epsilon_H^{}$ need to be of opposite sign. The reason is that now $A_t$ turns 
out to be negative at the GUT scale (cf.~the discussion in Section 
\ref{sec:running}). 

It is interesting to see how the mixing angles $\phi_Q$ and $\phi_L$ influence 
the nature of the LSP. $\phi_L$ determines the size of the stau parameters. Since
it is constrained by the tau Yukawa coupling 
it can only vary over a sizable range if $\tan\beta$ is large. 
The reason is that $A_\tau$ is generically large, leading to a charge-breaking 
minimum if $m_{L_3,E_3}$ are too small. Thus for $\tan\beta\sim 10$, 
$\phi_L$ is close to $\pi/2$ and the staus are rather heavy compared to the 
selectrons. For larger $\tan\beta$ (i.e.\ larger $y_\tau$), $\phi_L$ can be small and the 
$\tilde\tau_1$ can become the LSP, corresponding to the green points in 
Figs.~\ref{fig:BN-multiplot1} and \ref{fig:BN-multiplot2}. Note, however, that 
the stau LSP region is highly constrained by direct mass  bounds and $B$-physics,
and that a stable LSP is excluded by cosmology.
$\phi_L$ also has some effect on the selectron masses through RG evolution, but
this is much less pronounced.

The angle $\phi_Q$, on the other hand, determines the size of the stop and 
sbottom parameters. Through RG evolution it also influences the slepton masses, 
in particular $m_{\tilde e_R}$: larger  $\phi_Q$ leads to a larger $m_{D_3}$, 
which in turn decreases $m_{\tilde e_R}$. In Figs.~\ref{fig:BN-multiplot1} and 
\ref{fig:BN-multiplot2} one can see clearly that for increasing $\phi_Q$, the 
${\tilde e_R}$ eventually becomes the LSP. 
This behaviour can be understood easily from the $\rm U(1)_Y$ $D$-term contribution 
to the evolution of the scalar soft masses $m_i^2$~\cite{Martin:1993zk}. At one loop  
\be
  \frac{d}{dt}m_i^2 \sim \frac{6}{5}\frac{g_1^2 Y_i}{16\pi^2} S\,,
\ee
where $Y_i$ is the weak hypercharge and 
\be
  S=\left( m_{H_2}^2-m_{H_1}^2\right) + 
  {\rm Tr}\left(m_{Q}^2 - 2m_{U}^2 + m_{D}^2+m_{R}^2-m_{L}^2\right)
\ee  
with the trace running over generations. Since $S$ is an RG invariant, it simply causes 
a shift of the low-scale masses by  $\Delta m_i^2\approx -(0.052)\,Y_i\,S_{\rm GUT}$ 
\cite{Evans:2006sj} with respect to the values they would have had for $S\equiv0$. 
Here $S_{\rm GUT}$ is the value of $S$ at $M_{\rm GUT}$. 
For simplified boundary conditions, we had $S_{\rm GUT}=-m_{U_3}^2$.
With $Y_{e_R}=1$ and $Y_{e_L}=-1/2$, making $S_{\rm GUT}$ less negative 
obviously lowers $m_{\tilde e_R}$ and increases $m_{\tilde e_L}$ (note also that the 
effect for the left-chiral state is only half the size of that for the right-chiral one). 
Moreover, comparing $m_{\tilde e_R}\approx (0.39\,M_{1/2})^2 - 0.052\,S_{\rm GUT}$ 
to $m_{\tilde\chi^0_1}\approx 0.43\,M_{1/2}$, we understand why the ${\tilde e_R}$
eventually becomes the LSP.

\begin{figure}[htbp]\hspace*{-4mm}\begin{tabular}{rr}
   \includegraphics[width=7.9cm]{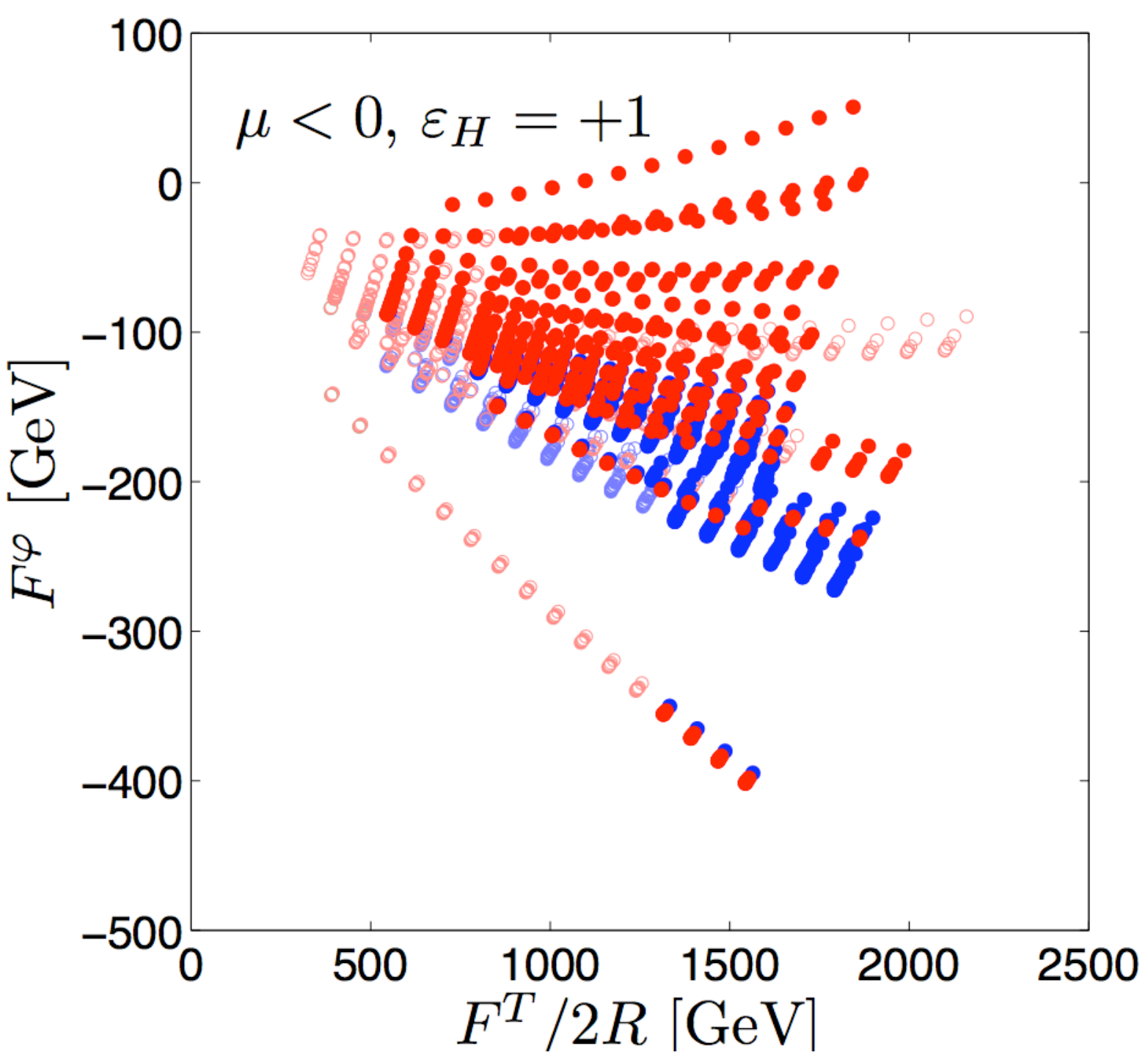}& 
   \includegraphics[width=7.9cm]{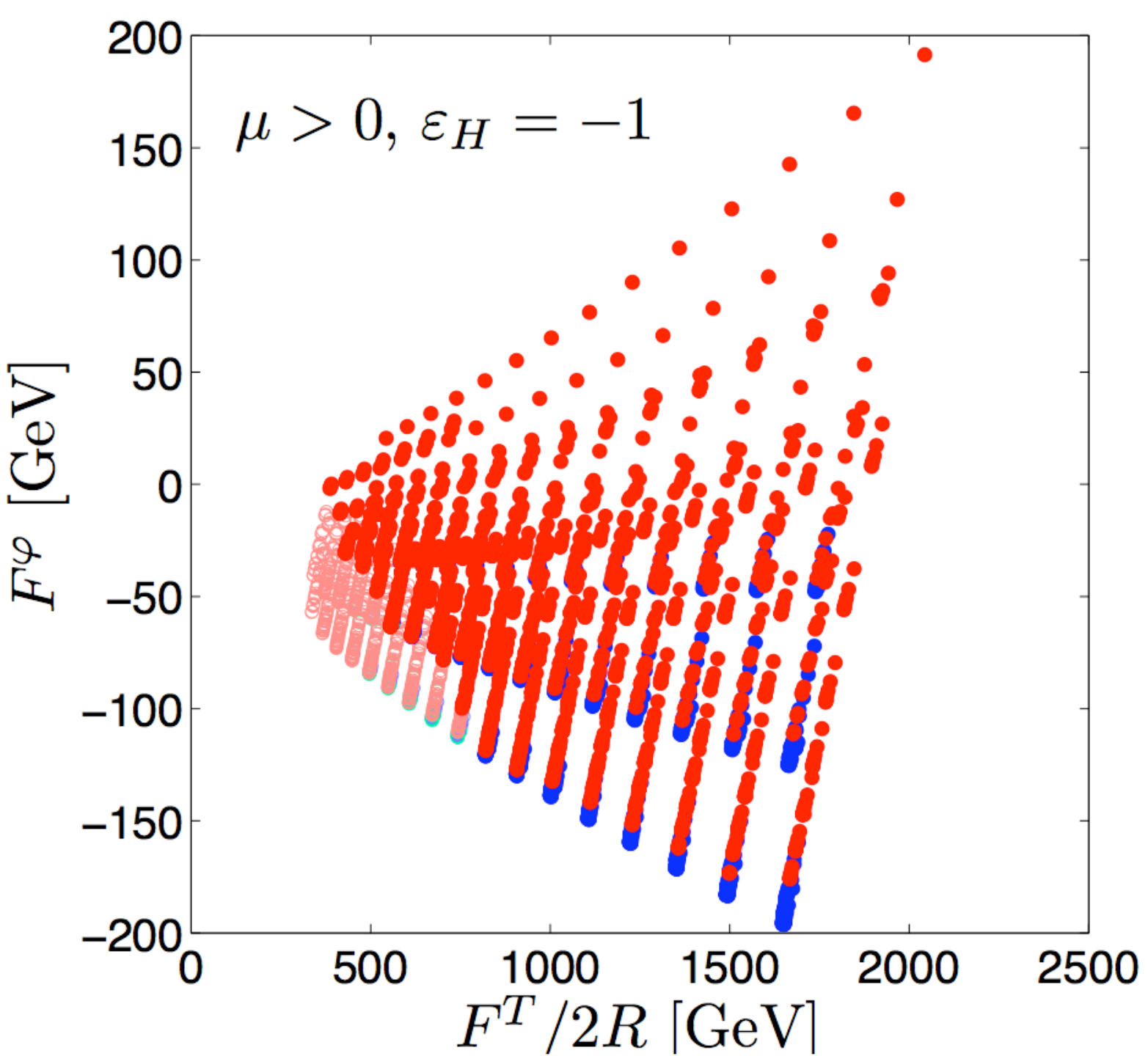}\\
   \includegraphics[width=7.7cm]{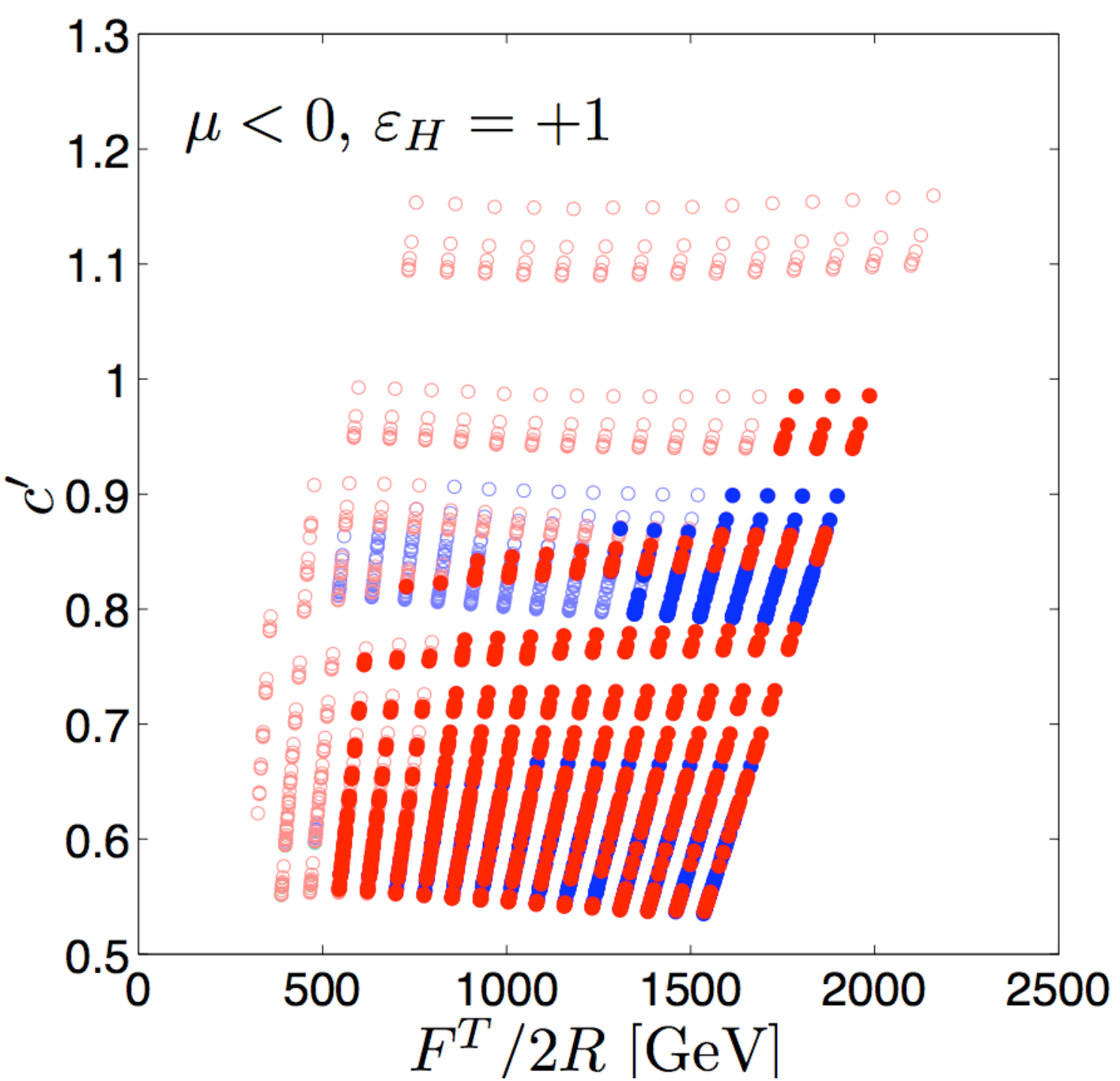}&
   \includegraphics[width=7.7cm]{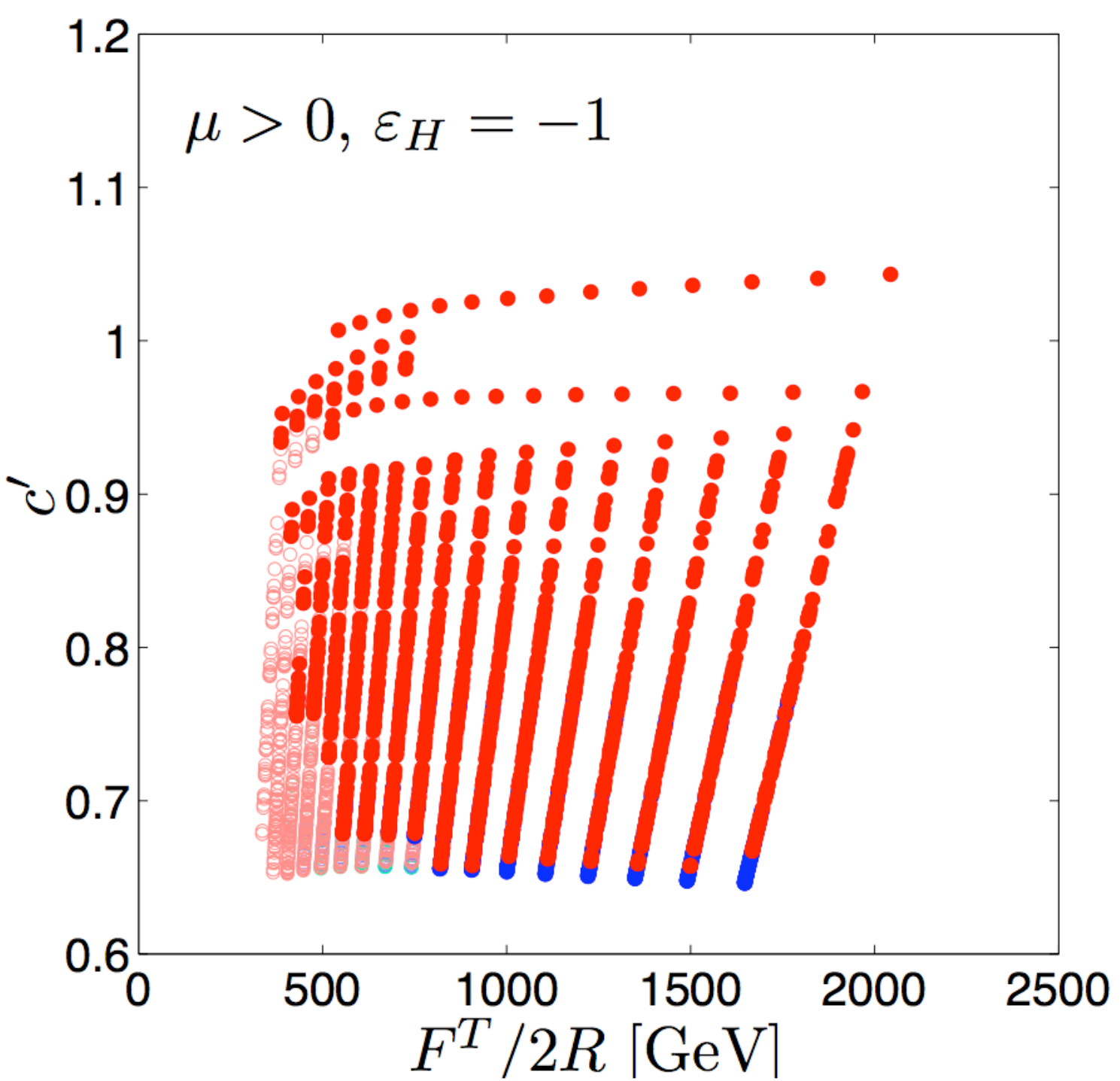}
   \end{tabular}
   \caption{Points of Figs.~\ref{fig:BN-multiplot1} and \ref{fig:BN-multiplot2}
   in the $F^T/2R$ vs.\ $F^\varphi$
   plane (top row) and in the $F^T/2R$ vs.\ $c'$ plane (bottom row). 
   The red, green and blue points have a neutralino, stau and selectron LSP, respectively.
   Open circles denote points excluded by B-physics constraints. 
   Points excluded by LEP are not shown.}
   \label{fig:BNfull-parspace}
\end{figure}

The projections onto the underlying model parameters $F^T/2R$, $F^\varphi$ and $c'$ 
are shown in Fig.~\ref{fig:BNfull-parspace}. 
Since here we need ${\rm sign}(\mu)=-\epsilon_H^{}$ 
to obtain a valid spectrum, 
$F^\varphi$ now turns out to be small and can even be zero. 
Contributions to the soft terms from anomaly mediation 
are therefore completely negligible.
Moreover, we find a somewhat smaller range for the $c'$ parameter, roughly 
$0.5\lesssim c' \lesssim1.2$, as compared to $0.5\lesssim c' \lesssim 3$ for simplified boundary conditions. 
The important point, however, is that $c'$ remains non-zero. We conclude    
that the Chern--Simons term is indeed essential to achieve correct EWSB.

\begin{figure}[t]\centering
   \includegraphics[width=8cm]{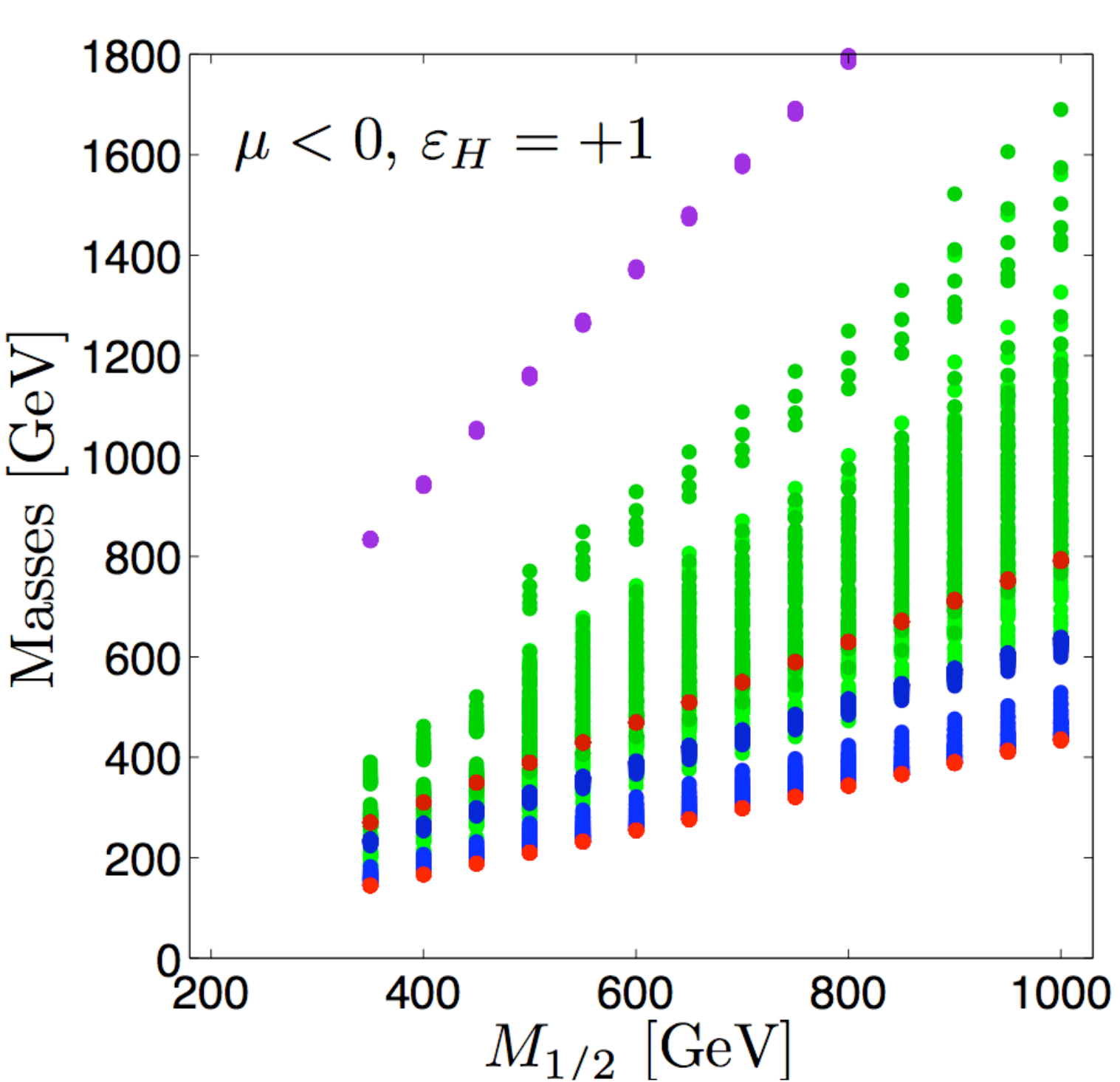} 
   \includegraphics[width=8cm]{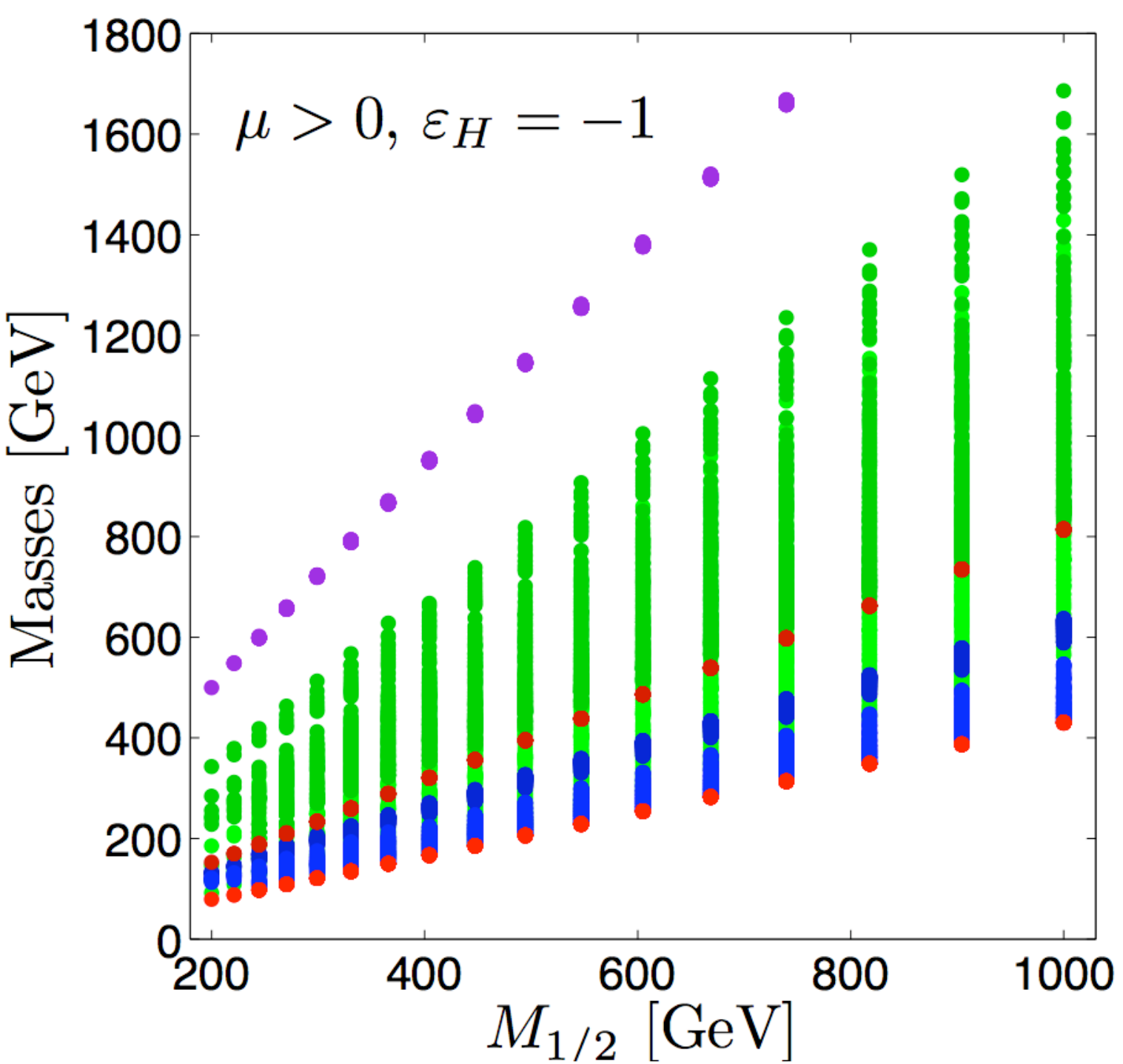} 
   \caption{Mass spectrum in the neutralino LSP region, passing LEP and B-physics constraints, 
   as a function of $M_{1/2}$. From bottom to top: 
   $\tilde\chi^0_1$ (red), $\tilde e_R$ (blue), $\tilde e_L$ (dark blue), 
   $\tilde\tau_1$ (green), $\tilde\chi^0_2$ (dark red), $\tilde\tau_2$ (dark green)
   and $\tilde g$ (purple).}
   \label{fig:BNfull-masses}
\end{figure}

Let us now turn to the implications for collider phenomenology. We again focus
on the neutralino LSP region. The mass spectrum in this region, taking into
account the constraints from LEP and from B-physics, is depicted in 
Fig.~\ref{fig:BNfull-masses}. As one can see, there is a definite mass ordering 
$m_{\tilde\chi^\pm_1}\simeq m_{\tilde\chi^0_2}>m_{\tilde e_L}>m_{\tilde e_R}>m_{\tilde\chi^0_1}$. 
The $\tilde\tau_2$ turns out to be heavier than the $\tilde\chi^0_2$, while the 
$\tilde\tau_1$ can be lighter than the $\tilde\chi^0_2$, and for small $\phi_L$ 
also lighter than the selectrons, cf.~the above discussion of the mixing-angle 
dependence. This gives a picture that is qualitatively similar to the simplified 
case discussed in Section~\ref{sec:simplenum}; the main difference lies in the 
masses and mass ratios of the sleptons. For the squarks, this effect of 
non-universality --- on the one hand the splitting of the third generation from 
the first and second generations due to non-zero $m_{Q_3,U_3,D_3}^2$, on the 
other hand the splitting of left- and right-chiral states due to non-zero $S$ 
--- is much less pronounced, because the running of the squark mass parameters 
is mainly driven by $M_3$. The squark and gluino masses are hence again about 
$m_{\tilde q}\approx m_{\tilde g}\approx(1.7-2.5)M_{1/2}$. 
The masses of the higgsino-like neutralinos and chargino are given by $|\mu|$ 
and lie above $m_{\tilde g}$. 

\begin{figure}[t]\centering
   \includegraphics[width=8cm]{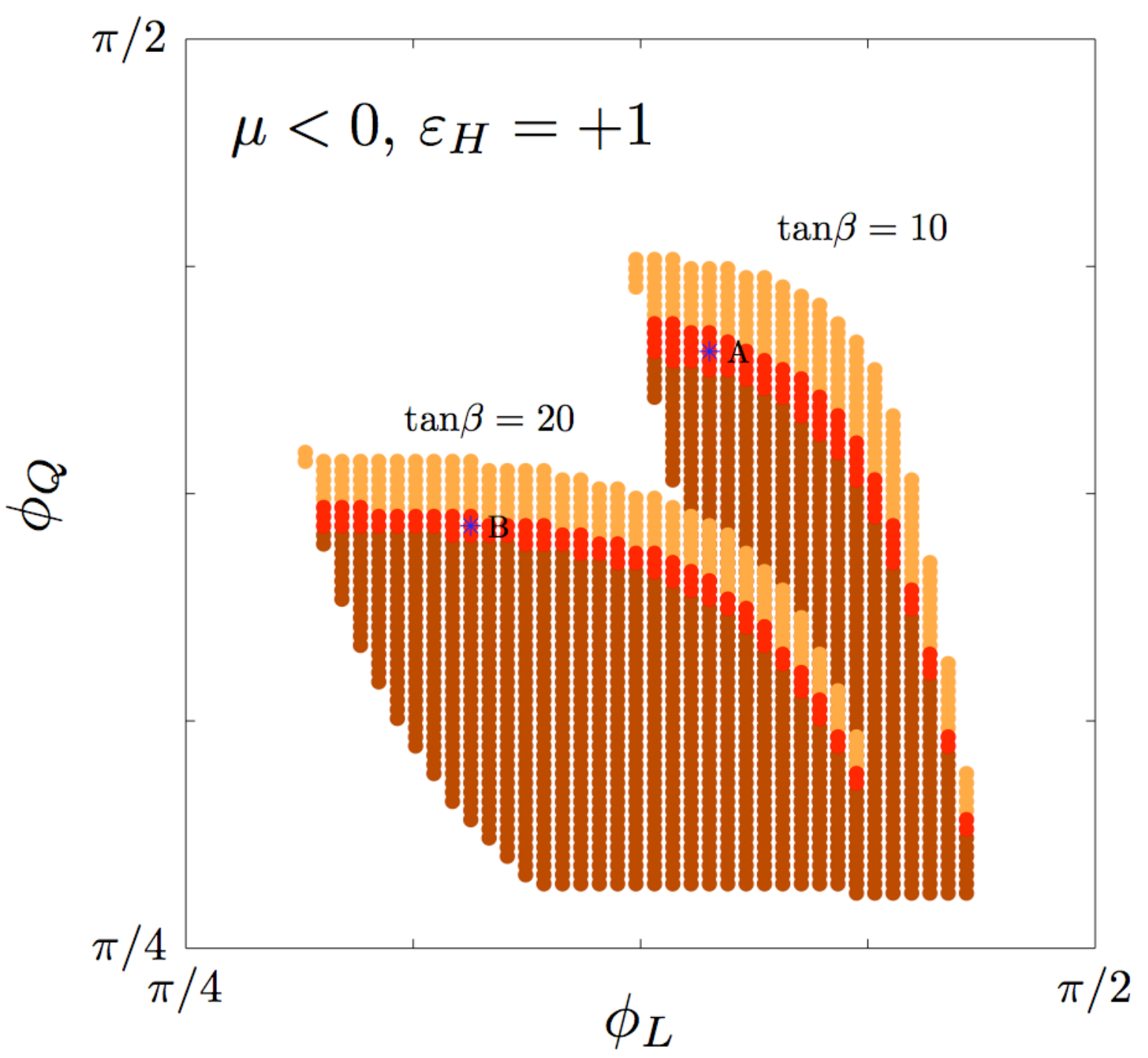} 
   \includegraphics[width=7.9cm]{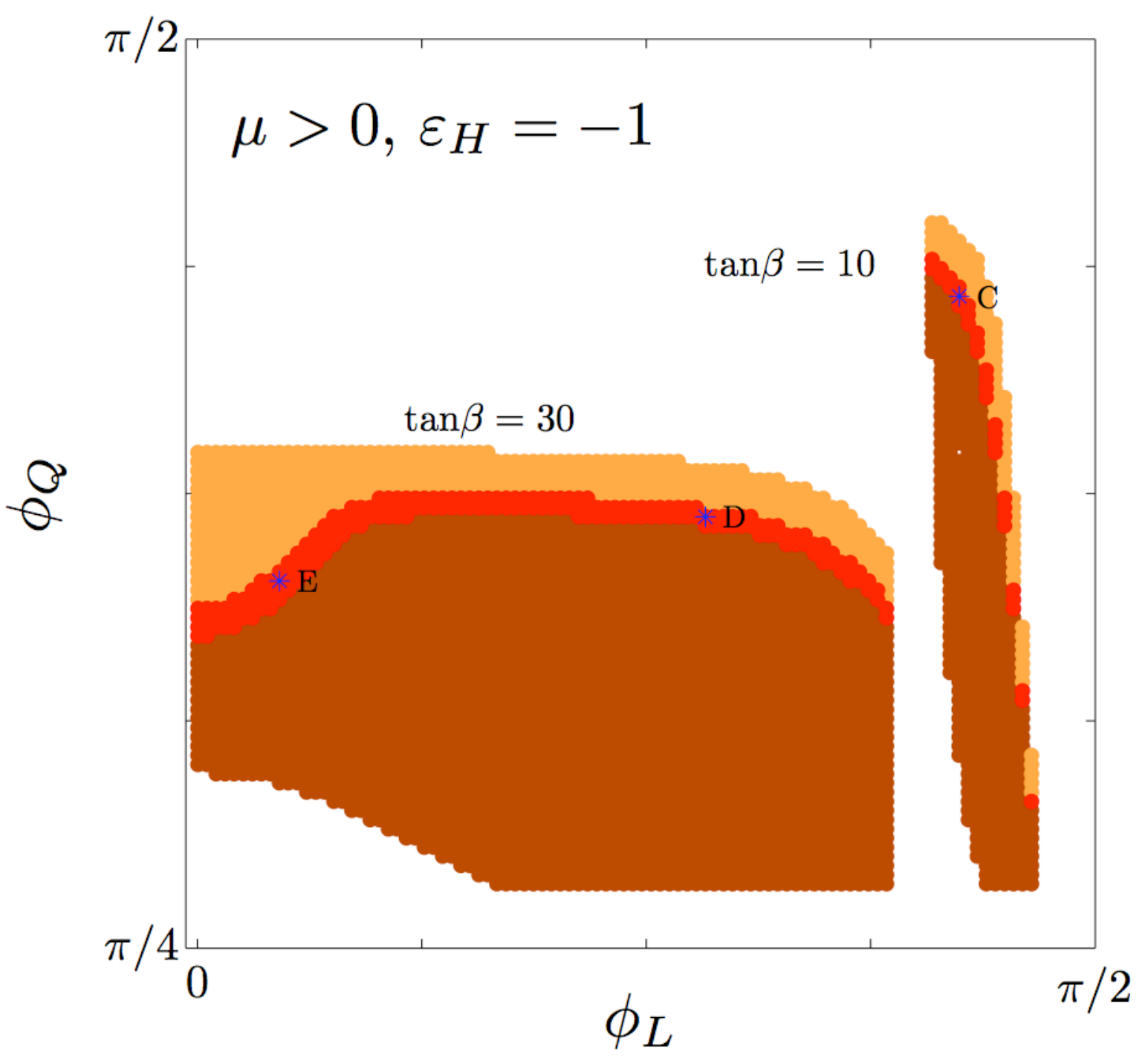} 
   \caption{Dependence of the neutralino relic density on the mixing angles $\phi_Q$ and $\phi_L$, 
   for $M_{1/2}=500$~GeV and various values of $\tan\beta$. 
   In the red bands, $\Omega h^2$ lies within $3\sigma$ of the WMAP5 observation,
   $0.0913<\Omega h^2<0.1285$. In the orange 
   regions, $\Omega h^2<0.0913$ is too low, while in the brown regions 
   $\Omega h^2>0.1285$ is too high. Also indicated are the sample points A--E.}
   \label{fig:BNfull-omega}
\end{figure}

It is also remarkable that now the neutralino relic density can vary over a large 
range, because of the extra parameters $\phi_Q$ and $\phi_L$.
This is illustrated in Fig.~\ref{fig:BNfull-omega} for the example of 
$M_{1/2}=500$~GeV and two values of $\tan\beta$ for each sign of $\mu$. For 
$\mu>0$, we take $\tan\beta=10$ and 30; for $\mu<0$, we take $\tan\beta=10$ 
and $20$ since higher values are too tightly constrained. The figure compares 
the neutralino relic density $\Omega h^2$, as a function of $\phi_Q$ and 
$\phi_L$, with the WMAP5 observation at $3\sigma$. In the orange regions 
$\Omega h^2$ is too low, which would require other constituents of dark matter in 
addition to the  neutralino. In the brown regions, on the other hand, 
$\Omega h^2$ is too high (at least within standard cosmology; it could be 
viable if there was, e.g., additional entropy production after freeze-out). 
The minimal and maximal values found are 
$\Omega h^2\simeq 6\times 10^{-3}$ and $0.9$, respectively. In the red band 
in between, however, $0.0913\le\Omega h^2\le 0.1285$ agrees within $3\sigma$ 
with the value measured by WMAP5. The reason is that here the mass difference 
between the LSP and NLSP (or co-NLSPs) is just right to make co-annihilation 
processes efficient enough, but not too efficient, to obtain 
$\Omega h^2\simeq 0.1$. To be precise, in the red bands of 
Fig.~\ref{fig:BNfull-omega} we typically have 
$\Delta m= m_{\tilde e_R}-m_{\tilde\chi^0_1}\simeq 7-10$~GeV. An exception is 
$\mu>0$, $\tan\beta=30$ and small $\phi_L$, where the $\tilde\tau_1$ becomes 
light and also contributes to co-annihilations, such that 
$\Delta m\approx 20$~GeV is needed; this leads to the red band bending down 
towards lower $\phi_Q$. Sample spectra of five representative points, indicated 
as points A--E in Fig.~\ref{fig:BNfull-omega}, are given in Tables~\ref{tab:parameters}  
and~\ref{tab:masses}. 
Table~\ref{tab:parameters} gives GUT and EWSB scale parameters, and 
Table~\ref{tab:masses} lists the resulting masses together with B-physics observables, 
the neutralino relic density and the neutralino--proton scattering cross section for direct 
detection. The possibility to tune 
the NLSP--LSP mass difference by adjusting $\phi_Q$ and $\phi_L$ and to obtain 
the correct relic density persists also for other values of $M_{1/2}$.

To summarize, the expected LHC phenomenology is as follows:
\begin{itemize} 
\item Squarks and gluinos with masses up to about 2~TeV will be 
  abundantly produced at the LHC, both as $\tilde q\tilde q$ or $\tilde g\tilde g$ pairs,  
  and in $\tilde q\tilde g$ associated production. They decay as $\tilde g\to q\tilde q_{R,L}^{}$, 
  $\tilde q_R^{} \to q\tilde\chi^0_1$ ($\sim 100\%$), $\tilde q_L^{}\to q'\tilde\chi^\pm_1$  
  ($\sim 65\%$) or $q\tilde\chi^0_2$ ($\sim 30\%$).
\item The decay $\tilde\chi^0_2 \to e^\pm \tilde e_L^\mp \to e^+ e^- \tilde\chi^0_1$  
         is always open and has a sizable branching ratio ($\sim$45\% for points A--D, 
         35\% for point E). This leads to a rather large 
         rate for the gold-plated SFOS dilepton signature.  
         In parts of the parameter space, also 
         $\tilde\chi^0_2 \to \tau^\pm \tilde\tau_1^\mp \to \tau^- \tau^+ \tilde\chi^0_1$ 
         can be kinematically allowed;
         c.f.\ point E, where it has 22\% branching ratio.
         Decays into $Z$, $h$, or $\tilde e_R$
         are negligible because the $\tilde\chi^0_2$ is almost a pure wino.
\item The decay of the $\tilde\chi^\pm_1$ always leads to a charged lepton, 
   $\tilde\chi^\pm_1\to \ell^\pm \tilde\nu_{\ell}/\nu_\ell \tilde \ell^\pm_L \to 
   \ell^\pm \nu_\ell \tilde\chi^0_1$, 
   giving rise to a large number of   
   events with jets plus one hard lepton plus 
   $E_T^{\rm miss}$. 
   If combined with $\tilde\chi^0_2\to ... \to l^+l^-\tilde\chi^0_1$ on the other side 
   of the event, it leads to the rather clean 
   trilepton signature (plus jets plus $E_T^{\rm miss}$).
\item The higgsino states $\tilde\chi^0_{3,4}$ and $\tilde\chi^\pm_2$ have masses around or 
above the gluino mass and are hence too heavy to be studied at the LHC.  
\end{itemize}

Overall the scenario resembles the mSUGRA/CMSSM case with small $m_0$, or the 
case of Higgs boson exempt no-scale supersymmetry (HENS) \cite{Evans:2006sj}.
An important difference are the sizeable third-generation high-scale soft
terms which our construction predicts.
Ways to distinguish between the different models include, e.g., the rate of 
leptonic events, which is expected to be higher in our scenario as compared to 
the mSUGRA case with the same $M_{1/2}$. Other distinctive features are the 
ratios of left- and right-chiral slepton masses and the non-universality of the 
third generation.\footnote{We leave a detailed study of characteristic mass 
ratios and model footprints for future work.} Note, however, that the 
$\tilde e_R$ does not couple to the wino-like $\tilde\chi^0_2$ and 
$\tilde\chi^\pm_1$ and hence does not appear in decay chains at the LHC. The 
$\tilde e_R$ is therefore best studied in $e^+e^-$ collisions, as are the staus 
if they are too heavy to be produced in $\tilde\chi^0_2$ decays.

Last but not least, a decisive test of GHU requires the precise measurement of 
the complete spectrum, including stops, sbottoms, heavy Higgs bosons and 
higgsinos, such that the SUSY Lagrangian parameters can be extracted and a 
bottom-up evolution along the lines of \cite{Blair:2000gy} performed. This can 
only be achieved at a (multi-)TeV $e^+e^-$ linear collider with a very good 
beam performance.

\section{Conclusions}\label{sec:conclusions}

We have investigated SUSY grand unified models with gauge-Higgs 
unification (GHU). A particularly interesting class of such models are 5d 
orbifold GUTs and heterotic string models which admit a 5d orbifold GUT 
limit. With the natural assumption of radion mediation, GHU models are 
quite predictive as far as the Higgs sector is concerned. The GUT-scale 
Higgs mass parameters are subject to the GHU relations, and are also tied 
to the gaugino mass. Despite these strong constraints, models of this type 
can be fully realistic, as we have shown. If the effects of a Chern--Simons 
term (which is generically present in 5d models) are taken into account, 
one finds regions in the parameter space which lead to proper electroweak 
symmetry breaking and satisfy the experimental bounds from direct Higgs and 
superpartner searches, rare decays and cosmology. We demonstrated this by 
using a variation of a 5d SU$(6)$ orbifold GUT model due to Burdman and 
Nomura as a concrete example. We gave detailed expressions for the soft 
SUSY breaking parameters in terms of the fundamental model data, including 
the Chern--Simons term. 

Using the high-scale relations between soft terms and estimates of running 
effects, we discussed qualitatively which parts of the parameter space 
might be promising. We then presented a detailed numerical analysis of the 
corresponding RGEs. This analysis was done in two parts, the first for a 
simplified model of the sfermion sector, and the second treating the 
relevant sfermion contributions properly as in the Burdman--Nomura model. 
The latter part of the analysis, while more realistic, is more involved 
because it depends on more parameters. In both cases we indeed find viable 
solutions to the RGEs, satisfying all present experimental constraints. A 
non-zero Chern--Simons term is essential to get a valid spectrum. 

We extracted some characteristic experimental signatures of this class of 
models, which will be tested at the LHC. In particular, selectrons are 
generically predicted to be lighter than the $\tilde\chi^0_2$, leading to a 
rather large rate for same-flavour opposite-sign dileptons over the whole 
parameter space. Higgsinos, on the other hand, are expected to be heavy, 
presumably beyond 
the reach of the LHC. Characteristic mass ratios could be tested in detail 
at a future $e^+e^-$ linear collider.

The LHC will have the potential to narrow down the allowed region in 
the parameter space of GHU models significantly, or to rule them out. 
This applies even more to a future linear collider. It would be worthwhile 
to study in detail how well the scenario discussed here could be 
reconstructed at the LHC and a linear $e^+e^-$ collider, thereby 
testing the GHU relation. To this end we proposed a set of benchmark 
points which may be useful for Monte Carlo simulations.

\section*{Acknowledgments}

We would like to thank B.~Allanach, D.~Cerde\~no, C.~Durnford, J.-L.~Kneur, M.~Ratz 
and K.~Schmidt-Hoberg for useful discussions. FB thanks the Physik-Department T30e 
of TU Munich for hospitality and support during the early stages of this project. 
This work is also supported by the French ANR project ToolsDMColl, BLAN07-2-194882.

\clearpage

\begin{table}\centering
\begin{tabular}{|c|r|r|r|r|r|}
\hline 
Point&
A&
B&
C&
D&
E\tabularnewline
\hline
\hline 
$M_{1/2}$&
500&
500&
500&
500&
500\tabularnewline
\hline 
tan$\beta$&
10&
20&
10&
30&
30\tabularnewline
\hline 
sign($\mu$)&
$-1$&
$-1$&
$+1$&
$+1$&
$+1$\tabularnewline
\hline 
$\varepsilon_{H}$&
$+1$&
$+1$&
$-1$&
$-1$&
$-1$\tabularnewline
\hline 
$\phi_{Q}$&
1.3011&
1.1503&
1.3486&
1.1582&
1.1027
\tabularnewline
\hline 
$\phi_{L}$&
1.2376&
1.0314&
1.3329&
0.8889&
0.1437\tabularnewline
\hline
\hline 
$c'$&
0.5712&
0.6118&
0.7577&
0.6686&
0.6811\tabularnewline
\hline 
$F^{T}/2R$&
785.6&
805.9&
878.9&
834.3&
840.9\tabularnewline
\hline 
$F^{\varphi}$&
$-107.7$&
$-120.6$&
$-38.9$&
$-114.6$&
$-109.9$\tabularnewline
\hline
\hline
\multicolumn{6}{|l|}{Parameters at $M_{\rm GUT}$} \tabularnewline
\hline
$\mu$&
$-1178.9$&
$-1232.4$&
$1296.6$&
$1283.2$&
$1291.0$\tabularnewline
\hline 
$B$&
$-344.0$&
$-381.3$&
$-298.0$&
$-393.2$&
$-390.2$\tabularnewline
\hline 
$m_{U_{3}}$&
564.8&
615.2&
622.9&
634.9&
657.3\tabularnewline
\hline 
$m_{D_{3}}$&
239.5&
371.2&
279.5&
413.9&
369.2\tabularnewline
\hline 
$m_{Q_{3}}$&
555.5&
592.4&
615.9&
613.5&
626.3\tabularnewline
\hline 
$A_{t}$&
$-812.6$&
$-793.7$&
$-980.0$&
$-842.2$&
$-823.1$\tabularnewline
\hline 
$A_{b}$&
$-932.3$&
$-923.6$&
$-1105.6$&
$-966.3$&
$-978.8$\tabularnewline
\hline 
$m_{R_{3}}$&
226.7&
304.9&
364.1&
432.6&
278.8\tabularnewline
\hline 
$m_{L_{3}}$&
161.8&
269.4&
225.2&
407.9&
278.5\tabularnewline
\hline 
$A_{\tau}$&
$-1055.8$&
$-1073.7$&
$-1216.8$&
$-1070.4$&
$-1149.6$\tabularnewline
\hline
\hline 
\multicolumn{6}{|l|}{Parameters at $M_{\rm EWSB}$} \tabularnewline
\hline
$\mu$&
$-1217.6$&
$-1238.8$&
$1342.0$&
$1275.3$&
$1282.9$ \tabularnewline
\hline 
$B$&
$-46.8$&
$-16.7$&
45.9&
11.4&
10.6\tabularnewline
\hline 
$m_{1}^2$&
554195&
405781&
598972&
428198&
401063\tabularnewline
\hline 
$m_{2}^{2}$&
$2022$&
$-2969$&
$2560$&
$-3429$&
$-3467$\tabularnewline
\hline
\end{tabular}
\caption{Parameters of sample points A--E in Fig.~\ref{fig:BNfull-omega}. 
Dimensionful quantities are in GeV, $B\equiv B\mu/\mu$.}
\label{tab:parameters}
\end{table}

\begin{table}\centering
\begin{tabular}{|c|r|r|r|r|r|}
\hline 
Point&
A\qquad&
B \qquad&
C \qquad&
D \qquad&
E \qquad\tabularnewline
\hline
\hline 
$m_{\tilde{\chi}_{1}^{0}}$&
210.1&
210.3&
208.1&
208.7&
208.7\tabularnewline
\hline 
$m_{\tilde{\chi}_{2}^{0}}$&
389.3&
389.5&
399.2&
400.4&
400.3\tabularnewline
\hline 
$m_{\tilde{\chi}_{3}^{0}}$&
1219.7&
1240.7&
1332.0&
1265.2&
1272.8\tabularnewline
\hline 
$m_{\tilde{\chi}_{4}^{0}}$&
1220.3&
1241.9&
1335.1&
1267.7&
1275.3\tabularnewline
\hline 
$m_{\tilde{\chi}_{1}^{\pm}}$&
389.3&
389.4&
399.2&
400.4&
400.3\tabularnewline
\hline 
$m_{\tilde{\chi}_{2}^{\pm}}$&
1222.8&
1244.1&
1335.4&
1268.4&
1276.0\tabularnewline
\hline
\hline 
$m_{\tilde{e}_{L}}$&
327.7&
328.1&
327.5&
326.8&
323.1\tabularnewline
\hline 
$m_{\tilde{e}_{R}}$&
218.6&
217.0&
216.6&
217.1&
228.3\tabularnewline
\hline 
$m_{\tilde{\tau}_{1}}$&
295.9&
322.0&
370.4&
387.7&
225.4\tabularnewline
\hline 
$m_{\tilde{\tau}_{2}}$&
372.6&
441.2&
438.0&
549.5&
457.0\tabularnewline
\hline 
$m_{\tilde{\nu}_{e}}$&
318.4&
318.3&
318.1&
317.3&
313.4\tabularnewline
\hline 
$m_{\tilde{\nu}_{\tau}}$&
354.3&
408.0&
386.2&
499.0&
398.4\tabularnewline
\hline
\hline 
$m_{\tilde{u}_{L}}$&
1046.1&
1045.8&
1042.1&
1041.8&
1041.9\tabularnewline
\hline 
$m_{\tilde{u}_{R}}$&
1003.5&
1003.2&
1000.0&
999.6&
997.6\tabularnewline
\hline 
$m_{\tilde{d}_{L}}$&
1049.0&
1048.7&
1045.1&
1044.8&
1044.9\tabularnewline
\hline 
$m_{\tilde{d}_{R}}$&
1005.9&
1005.6&
1001.9&
1001.8&
1002.3\tabularnewline
\hline 
$m_{\tilde{t}_{1}}$&
948.0&
971.1&
955.7&
971.4&
983.5\tabularnewline
\hline 
$m_{\tilde{t}_{2}}$&
1147.0&
1155.0&
1187.9&
1167.6&
1175.2\tabularnewline
\hline 
$m_{\tilde{b}_{1}}$&
1022.3&
1016.7&
1029.9&
1021.7&
1007.9\tabularnewline
\hline 
$m_{\tilde{b}_{2}}$&
1108.4&
1119.6&
1137.6&
1130.6&
1135.5\tabularnewline
\hline 
$m_{\tilde{g}}$&
1155.5&
1156.5&
1154.2&
1154.7&
1155.0\tabularnewline
\hline
\hline  
$m_{h}$&
115.0&
116.4&
117.2&
117.3&
116.8\tabularnewline
\hline
$m_{H}$&
762.4&
658.5&
770.7&
637.4&
635.9\tabularnewline
\hline 
$m_{A}$&
761.6&
658.5&
770.8&
637.6&
635.9\tabularnewline
\hline 
$m_{H^{\pm}}$&
766.7&
663.9&
775.1&
642.9&
641.0\tabularnewline
\hline
\hline 
BR$(b\rightarrow s\gamma)$&
$3.70\times10^{-4}$&
$4.16\times10^{-4}$&
$3.20\times10^{-4}$&
$2.89\times10^{-4}$&
$2.91\times10^{-4}$ \tabularnewline
\hline 
BR$(B_{s}\rightarrow\mu^{+}\mu^{-})$&
$2.89\times10^{-9}$&
$2.10\times10^{-9}$&
$3.06\times10^{-9}$&
$6.76\times10^{-9}$&
$6.68\times10^{-9}$ \tabularnewline
\hline 
$\Omega h^{2}$&
0.110&
0.108&
0.110&
0.108&
0.106\tabularnewline
\hline 
$\sigma(\tilde{\chi}p)^{\rm SI}$ {[}pb{]}&
$2.92\times10^{-11}$&
$1.39\times10^{-10}$&
$1.01\times10^{-10}$&
$2.90\times10^{-10}$&
$2.89\times10^{-10}$\tabularnewline
\hline
\end{tabular}
\caption{Masses (in GeV), B-physics observables, relic density and 
spin-independent neutralino--proton scattering cross section for points A--E.}
\label{tab:masses}
\end{table}

\clearpage

\end{document}